\documentclass[a4paper,fleqn,usenatbib]{mnras}

\usepackage{mathptmx}

\usepackage[T1]{fontenc}
\usepackage{ae,aecompl}

\usepackage{graphicx}	
\usepackage{amsmath}	
\usepackage{amssymb}	
\usepackage{ifxetex}

\newcommand\RP{(R$^{\prime}$)}

\newcommand\urs{($\underline{\rm{r}}$s)}
\newcommand\uAB{$\underline{\rm{A}}$B}
\newcommand\AuB{A$\underline{\rm{B}}$}
\newcommand\rus{(r$\underline{\rm{s}}$)}

\newcommand\rul{(r$\underline{\rm{l}}$)}

\newcommand\uurl{($\underline{\rm{r}}$l)}

\title[Optical Morphologies of 719 Isolated Galaxies]{A Comprehensive Examination of the Optical Morphologies of 719 Isolated Galaxies in the AMIGA Sample}

\author[R. J. Buta et al.]{Ronald J. Buta$^1$, Lourdes Verdes-Montenegro$^2$, Ancor Damas-Segovia$^2$,
\newauthor{Michael Jones$^2$, Javier Blasco$^2$, Mirian Fern\'andez-Lorenzo$^2$, Susana Sanchez$^2$,}
\newauthor{Julian Garrido$^2$, Pablo Ramirez-Moreta$^2$, and J. Sulentic$^2$}
\\
$^1$Department of Physics \& Astronomy,University of Alabama, Box
870324, Tuscaloosa, AL 35487
\\
$^2$Department Astronomia Extragal\'actica, Instituto Astrofisica Andalucia s/n,
18008 Granada, Spain
}
\date{Accepted XXX. Received YYY; in original form ZZZ}

\pubyear{2017}

\begin{document}
\label{firstpage}
\pagerange{\pageref{firstpage}--\pageref{lastpage}}
\maketitle

\begin{abstract}
Using images from Sloan Digital Sky Survey Data Release 8, we have
re-examined the morphology of 719 galaxies from the Analysis of the
interstellar Medium in Isolated GAlaxies (AMIGA) project, a sample
consisting of the most isolated galaxies that have yet been identified.
The goal is to further improve the classifications of these galaxies by
examining them in the context of the Comprehensive de Vaucouleurs
revised Hubble-Sandage (CVRHS) system, which includes recognition of
features that go beyond the original de Vaucouleurs point of view. Our
results confirm previous findings that isolated galaxies are found
across the complete revised Hubble sequence, with intermediate to
late-type (Sb-Sc) spirals being relatively more common. Elmegreen Arm
Classifications are also presented, and show that more than 50\% of the
514 spirals in the sample for which an arm class could be judged are
grand design (AC 8,9,12). The visual bar fraction for the sample is
$\approx$50\%, but only 16\% are classified as strongly-barred (SB).
The dominant family classification is SA (nonbarred), the dominant
inner variety classification is (s) (pure spiral), and the dominant
outer variety classification is no outer ring, pseudoring, or lens. The
Kolmogorov-Smirnov test is used to check for potential biases in the
morphological interpretations, and for any possible relation between
rings, bars, and arm classes with local environment and far-infrared
excess. The connection between morphology and stellar mass is also
examined for a subset of the sample.

\end{abstract}



\section{Introduction}

From a morphological point of view, isolated galaxies are more
important than galaxies where environmental effects play a strong role
in galactic evolution. If the process of baryonic matter collecting
into a seed cold dark matter halo can form an individual galaxy (e.g.,
Firmani \& Avila-Reese 2003), then if that galaxy is sufficiently
isolated, its morphology would evolve according to the characteristics
it was endowed with at birth. If internal perturbations like bars and
spirals can spontaneously develop from an initially featureless disk,
then the evolution of an isolated galaxy will be strongly influenced by
how effectively these perturbations can drive the migrations of gas
clouds and stars over time. Isolated galaxies let us see the products
of ``nature" in galaxy structure. After the rapid formation process is
largely completed, a slow secular evolution would take over to modify a
galaxy's morphology (e.g., Kormendy 2012, 2014).

Because galaxies tend to be gregarious, compiling a truly isolated
galaxy sample is challenging. The best catalogue to date, and the one
studied in the most detail over a wide range of wavelengths, is the
Catalogue of Isolated Galaxies (CIG; Karachentseva 1973). The CIG was
based on inspection of Palomar Sky Survey charts, and includes 1050
entries selected using an isolation criterion that attempts to exclude
galaxies with similar-sized companions. Specifically, to get into
the CIG, a galaxy of angular diameter $D$ had to have no companion of
angular diameter $d$ between 1/4$D$ and 4$D$ that lies within an
angular separation of 20$d$. Small companions are not necessarily
ruled out, but the criterion assumes that, if a companion is small, it
is probably a background object.

A logical followup to Karachentseva's work is to re-examine the CIG
sample with better image material than the catalogue was based upon,
and to collect objective, wide-ranging information on the observed
properties of the galaxies. This inspired the Analysis of the
interstellar Medium of Isolated GAlaxies (AMIGA) project  begun by
Verdes-Montenegro et al. (2005), who also augmented the original
isolation criterion to make an isolated galaxy one which has not
experienced a major encounter in at least the last $\approx$3 Gyr.
This assumes a typical value of $D$ is 25 kpc and a typical field
velocity of 150 km s$^{-1}$. With such a criterion, we cannot say
what the average merger activity over the lifetime of the AMIGA sample
galaxies has been, only that there has been no activity for at least 3
Gyr.

AMIGA has refined the CIG and given it a multiwavelength
characterization. The AMIGA CIG can be described as a vetted or
value-added catalogue, based upon the original CIG. The degree of
isolation was re-evaluated and quantified for each galaxy in terms of
both the local number density of neighbours and tidal strength (Verley
et al. 2007a,b; Argudo et al.  2011; Argudo-Fern\'andez et al 2013).
AMIGA has clearly established the parameters expected to be enhanced by
interactions, such as level of optical asymmetry, clumpiness and
concentration (Durbala et al 2009), MIR/FIR luminosity (Lisenfeld et
al. 2007), radio continuum emission (Leon et al. 2008), radio-excess
above the radio-FIR correlation (0\%; Sabater et al. 2008; Sabater et al.
2010), AGN rate (22\%; Sabater et al 2012), HI asymmetry (Espada et
al.  2011), and the molecular gas content (Lisenfeld et al. 2011). All
of these physical characteristics are found at lower levels in isolated
galaxies than in any other sample, even compared with field galaxies,
while colours are redder\footnote{This has been interpreted by
Fern\'andez-Lorenzo et al. (2012) as due to a ''more passive star
formation in very isolated galaxies."} and disks larger
(Fern\'andez-Lorenzo et al 2012, 2013). 

Sulentic et al. (2006) used the deeper and finer grained prints of the
Palomar II sky survey to improve upon the classifications for CIG
galaxies using the system of the Third Reference Catalogue of Bright
Galaxies (RC3, de Vaucouleurs et al. 1991). This showed that the most
common types of galaxies in the CIG are Sb-Sc spirals, and that
early-type galaxies (ellipticals and S0s) are a non-negligible fraction
of the sample. Because of the limitations of these charts (small image
scale, frequent overexposure of the central regions; nonlinear
intensity scale), statistics of other morphological features (such as
bars, rings, lenses, ovals, etc.) were more difficult to quantify
reliably.

Digital imaging can provide the best information on isolated galaxy
morphologies.  Revised classifications of 843 CIG galaxies based mainly
on Sloan Digital Survey (SDSS; Gunn et al. 1998; York et al. 2000)
images were presented by Fern\'andez-Lorenzo et al (2012) for 843 CIG
galaxies with heliocentric radial velocity $V_{hel}$ $>$ 1000 km
s$^{-1}$. Less certain types for 191 additional CIG galaxies were
included.  Uncertainties in the revised morphological types depend on
image quality and angular resolution compared to galaxy size. A general
shift of $\Delta(T)$ = 0.2 in the revised numerical stage indices of
Fern\'andez-Lorenzo et al (2012) was found with respect to Sulentic et
al. (2006), likely due to the higher quality of the CCD images compared
to sky survey images.

One of the most extensive quantitative analyses of SDSS images of
isolated galaxies was made by Durbala et al. (2008, 2009). In addition
to also judging morphological types, these authors used both parametric
and non-parametric approaches to quantify the structure of about 100
AMIGA spirals of types Sb to Sc. In the parametric approach,
two-dimensional decompositions were used to derive Sersic indices, disk
radial scalelengths, and bulge-to-total luminosity ratios.
Non-parametric quantities like those provided by the
Concentration-Asymmetry-Clumpiness (CAS) system, relative Fourier
intensity amplitudes, and bar and spiral torque strengths were also
derived. Durbala et al. (2008, 2009) concluded that isolated galaxies
are less clumpy, less concentrated, more symmetric, and may have larger
bars than in samples of less isolated galaxies. In addition, these
authors found that  most AMIGA spirals host pseudobulges rather than
classical bulges (see also Fern\'andez-Lorenzo et al 2014). Other
commonly-used non-parametric quantities are discussed by Andrae,
Jahnke, and Melchior (2011).

In this paper, we use images from SDSS Data Release 8 (Aihara
et al. 2011) to examine the morphology of 719 AMIGA galaxies in the
Comprehensive de Vaucouleurs revised Hubble-Sandage (CVRHS)
classification system (Buta et al. 2007, 2015). The reasons for doing
this are: to improve upon our knowledge of likely nurture morphology;
to evaluate previous studies of CIG galaxy morphology; and to examine
statistics of bars, rings, lenses, and other features that have
received only partial attention in previous studies of isolated galaxy
morphology. Sections 2-4 describe the selection of the sample, the
procedure used to classify the galaxies, and an analysis of the
internal consistency of the classifications. Section 5 compares the new
classifications with previously published types from other sources,
while section 6 examines some of the basic morphological
characteristics of of the sample.  Section 7 examines correlations
between morphology and stellar mass that are present in the isolated
galaxy sample. Finally, in section 8, we look for possible correlations
between the CVRHS morphologies of our 719 isolated galaxies and other
parameters from the AMIGA database in order to explore potential
mechanisms that give rise to the different morphological features
(i.e., inner rings, outer rings, bars) classified in this sample. This
study provides us with a tool to find tendencies that could arise from
observational biases due to the limitations of the optical observations
or from real dependencies of morphological aspects on the evolutionary
stage of isolated galaxies.

\section{Data and Sample}

\begin{figure}
\includegraphics[width=\columnwidth]{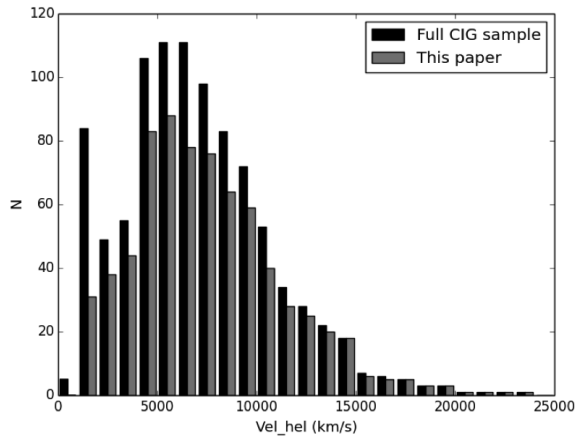}
\caption{Distribution of radial velocities for the full CIG sample and
our subset of 719 galaxies. This shows that both samples cover
about the same redshift range (0.005 to 0.080), meaning that our capacity to separate
details in the morphology of the galaxies will be roughly equivalent.}
\label{fig:radvels}
\end{figure}

Images from SDSS DR8 are available for 843 CIG galaxies. Those with
$V_{hel}$ $<$ 1500 km s$^{-1}$ were excluded from our analysis because
they are too nearby for a proper determination of isolation; also, some
of the 843 do not yet have a radial velocity available. This leaves $N$
= 719 CIG galaxies for our study. We show in Figure~\ref{fig:radvels}
the velocity distribution of the selected sample as compared to the
full CIG sample. The images are comparatively deep, but seeing quality
is variable in the dataset, adding some uncertainty to the
classifications. Many of the galaxies are also distant enough to have
not been included in RC3, and thus resolution is also an
issue.\footnote{For a detailed discussion of resolution effects on
CVRHS classifications of inner, outer, and nuclear varieties with SDSS
images, see section 3 of Buta 2017a.} The redshift range is $z$ =
0.005 to 0.080. The ranges of other parameters are given in
Table~\ref{tab:props}.

\begin{figure}
\includegraphics[width=\columnwidth]{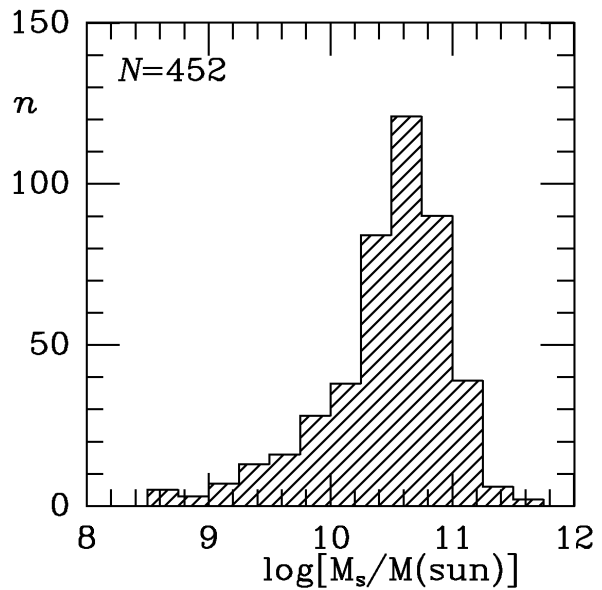}
\caption{Distribution of stellar masses for 452 (63\%) of our sample of
719 CIG galaxies. The mass estimates are from Fern\'andez et al.
(2013).}
\label{fig:masses}
\end{figure}

Figure~\ref{fig:masses} shows the logarithmic distribution of
stellar masses $M_s$ for 63\% of our selected sample. Although the
range of $log{M_s\over M_{\odot}}$ is 8.3 to 11.4, the median value is
10.57, meaning that dwarfs are not a major part of our sample. Assuming
the 452 galaxies in Figure~\ref{fig:masses} are representative of the
full sample, most of our sample galaxies are in the stellar mass range
close to the ``knee" in the stellar mass function, where the
characteristic mass is $log{M^*\over M_{\odot}}$=10.65 (Baldry,
Glazebrook, and Driver 2008). Also, the bulk of our sample galaxies lie
near the higher mass part of the blue cloud (Kelvin et al. 2018). The
correlation between stellar mass and specific aspects of morphology is
described further in section 7.

\begin{table}
\centering
\setcounter{table}{0}
\caption{Sample properties (Argudo-Hern\'andez et al. 2013, 2015)}
\label{tab:props}
\begin{tabular}{ll}
\hline
Parameter & Range \\
1 & 2 \\
\hline
redshift           & 0.005 $<$ $z$ $<$ 0.080 \\
$r$-band magnitude & 11 $<$ $m_r$ $<$ 15.7 \\
color index        & 0.2 $<$ $g-r$ $<$ 1.2 \\
linear diameter    & 1.1 kpc $<$ $D$ $<$ 23.2 kpc \\
stellar mass       & 8.32 $<$ $log{M_s\over M_{\odot}}$ $<$ 11.42 \\
\hline
\end{tabular}
\end{table}

\section{CVRHS Morphology and Isolated Galaxies}

In Buta et al. (2015), CVRHS morphology is described and applied to
mid-infrared (3.6$\mu$m) images from the Spitzer Survey of Stellar
Structure in Galaxies (S$^4$G; Sheth et al. 2010). Although it was
shown that the CVRHS system can be applied effectively in the mid-IR,
the historical basis for the system is the $B$-band, the effective
wavelength of which is only 0.44$\mu$m. Thus, to minimize systematic
effects, it is best to use images obtained with a $B$-band filter or a
filter close to the $B$-band.  In the case of SDSS images, the filter
closest to $B$ is the $g$-band at 0.477$\mu$m. We base our new
classifications mainly on logarithmic, sky-subtracted $g$-band images
in relative units of magnitudes per square arcsecond.\footnote{
These images were already pre-processed (i.e., background-subtracted,
field-selected) for other studies (as in, for example, Durbala et al.
2008, 2009).} This places the images in the ``classification ready"
mode of the de Vaucouleurs Atlas of Galaxies (Buta, Corwin, and Odewahn
2007=deVA).  In principle, SDSS colour images can also be used for
CVRHS classification, but these lack the dynamic range needed to see
morphology in the bright centers of some galaxies and were not used for
the classifications in this paper. Images in what the deVA refers to as
``atlas units" have a broader dynamic range. The range used for the
classifications from the AMIGA sample of images was approximately the
same for all of the images. For calibrated SDSS $g$-band images, the
range was 15.0-27.0 mag arcsec$^{-2}$.

The CVRHS system is a version of the de Vaucouleurs revised
Hubble-Sandage (VRHS) system (de Vaucouleurs 1959) that takes into
account more details of galaxy morphology that are of interest at the
present time but at the same time preserves the main features of the
original system. These details include lenses (Kormendy 1979), outer
resonant subclass rings and pseudorings (Buta \& Crocker
1991; Buta 1995, 2017b), ansae bars (Danby 1965; Martinez-Valpuesta et al.
2007; Buta 2012, 2013), nuclear rings (Burbidge \& Burbidge 1960; Buta
\& Crocker 1993; Comer\'on et al. 2010), nuclear bars (de Vaucouleurs
1975; Buta \& Crocker 1993; Erwin 2004), thick disks (Burstein 1979;
Comeron et al.  2011), inclined (extraplanar) rings (Schweizer, Whitmore, and Rubin
1983), disky and boxy ellipticals (Kormendy \& Bender 1996), spheroidal
galaxies (Kormendy 2012), nuclear lenses (Buta \& Combes 1996;
Laurikainen et al. 2013), and barlenses (Laurikainen et al. 2013).
Table 1 of Buta et al. (2015) summarizes all of the notation and
features of CVRHS galaxy classification.

In addition to CVRHS classifications, we also present Arm Classes (ACs)
for 514 of the 597 spiral galaxies in the sample, guided by Table 1
of Elmegreen \& Elmegreen (1987). Arm Classes are based on symmetry and
extent of the spiral arms in a galaxy, but can be uncertain or
indeterminate when the inclination is high, as is the case for many of
the 83 unclassified cases. Other reasons arm classes could
only be estimated for 86\% of the sample spiral galaxies are because
either the stage is too early ($T \leq$ 0), the stage is too late ($T
\geq$ 9), or the object is poorly resolved.  Arm classes are useful in
the context of the AMIGA sample because of the general view that spiral
arms are a transient phenomenon requiring a ``trigger," like a bar or a
close companion (e.g., Kormendy \& Norman 1979). Genuine examples of
isolated grand design spirals would imply that strong spiral arms could
nevertheless arise spontaneously within a disk without the presence of
a major companion.

\section{Procedure}

\begin{figure*}
\includegraphics[width=\textwidth]{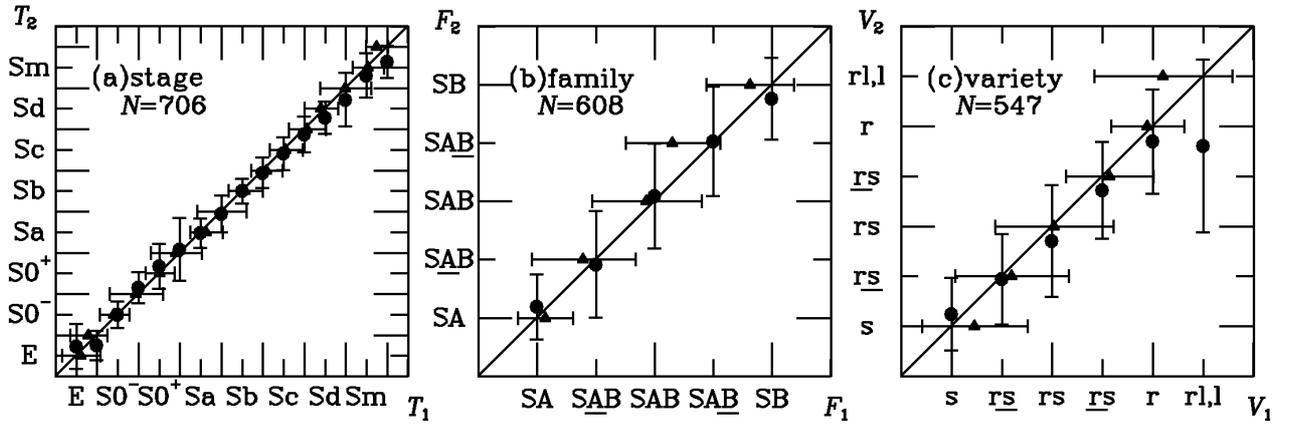}
\caption{Comparison of Phase 1 and 2 classifications (Table~\ref{tab:phase12}) with respect to stage, family, and variety.
}
\label{fig:p12}
\end{figure*}

Following Buta et al. (2015), the 719 AMIGA galaxies were classified
from the prepared ``classification ready" images by RB in two
phases separated by more than 6 months. The main reason for this is to
check the internal consistency of the full classifications.
Table~\ref{tab:phase12} includes the classifications from each phase,
and Figure~\ref{fig:p12} shows comparisons of stage, family, and
variety classifications between the two phases. In general, the
agreement between the two phases is very good, and for the purposes of
the remaining analysis, we take an unweighted average of the two phases
using numerical indices listed in Table 3 of Buta et al. (2015). These
unweighted average classifications are listed in
Table~\ref{tab:catalog}.

 \begin{table*}
 \centering
 \caption{Phase 1 and 2 CVRHS Classifications for 719 AMIGA Galaxies. Col. 1: number in Karachentseva (1973) catalogue; col. 2: number in Principal Galaxy Catalogue (Paturel et al.  1989); col. 3: full phase 1 classification; col. 4: full phase 2 classification. (The full table will be made available online.)}
 \label{tab:phase12}
 \begin{tabular}{llll}
 \hline
 Galaxy & PGC & Phase 1 type & Phase 2                   type \\
 1 & 2 & 3 & 4 \\
 \hline
{\bf $\rightarrow$ RA: 0$^h$ $\leftarrow$} & & & \\                                                                                                                                                                                         
CIG0001         &   205 & SA(r$\underline{\rm s}$)bc                                                                           & SA(rs)bc                                                                                             \\    
CIG0002         &   223 & SB(rs)cd                                                                                             & SB(rs)cd                                                                                             \\    
CIG0004         &   279 & SA(rs)c sp                                                                                           & SA(rs)c                                                                                              \\    
CIG0005         &   602 & SA(s)b:                                                                                              & SA(rs)b:                                                                                             \\    
CIG0006         &   652 & SAB(s)m/RG? pec                                                                                    & SB(s)m pec or Pec (merger)                                                                           \\    
CIG0007         &   793 & SA$\underline{\rm B}$(rs)b                                                                           & SAB(rs)ab                                                                                            \\    
CIG0008         &   833 & SA(s)cd                                                                                              & SA(s)cd                                                                                              \\    
CIG0009         &   859 & SA(s)d:                                                                                              & SAc pec                                                                                              \\    
CIG0011         &   963 & S$\underline{\rm A}$B(r$\underline{\rm s}$)d                                                         & SA(s)cd                                                                                              \\    
CIG0012         &  1056 & SAB(s)m sp pec                                                                                       & Sb: spw pec                                                                                          \\    
 \hline
 \end{tabular}
 \end{table*}

As in Buta et al. (2015), the averaging of two catalogues in this
manner leads to extensive use of the de Vaucouleurs (1963) underline
notation. This notation is meant to emphasize a particular part of
a combined characteristic. For example, the family classification
S$\underline{\rm A}$B implies a galaxy with only a trace of a bar; i.e.,
the galaxy is mostly nonbarred, while SA$\underline{\rm B}$ implies a
galaxy with a clear but not strong bar, i. e., the galaxy is mostly
barred. An inner variety of ($\underline{\rm r}$s) is a mostly closed
inner pseudoring while an inner variety of (r$\underline{\rm s}$) is a
mostly open inner pseudoring. The underline classifications for family
and inner variety are well-enough defined to be applied directly, i.e.,
do not appear just because the final catalogue is an average of two
phases. In principle, underline stages [like S$\underline{\rm a}$b
(more Sa than Sb) or Sc$\underline{\rm d}$ (more Sd than Sc)] could
also be applied directly. However, this is more difficult for stages,
and underline stages only appear in average multi-phase
classifications, mainly to preserve information.

 \begin{table*}
 \centering
 \caption{Mean CVRHS Classifications for 719 AMIGA Galaxies. Col. 1: number in Karachentseva (1973) catalogue; col. 2: number in Principal Galaxy Catalogue (Paturel et al. 1989); col. 3; mean numerical stage index; col. 4: Elmegreen arm class; col. 5: mean full classification; col. 6: notes (no abbreviations). (The full table will be made available online.)}
 \label{tab:catalog}
 \begin{tabular}{llrrll}
 \hline
 Galaxy & PGC & $<T>$ & AC & $<$Type$>$  &               notes \\
 1 & 2 & 3 & 4 & 5 & 6 \\
 \hline
{\bf $\rightarrow$ RA: 0$^h$ $\leftarrow$}  & & & & & \\                                                                                                                                                                                    
CIG0001         &   205 &    4.0 &  9  & SA(rs)bc              & UGC 5; excellent case   \\                                                                                                                                                 
CIG0002         &   223 &    6.0 &  5  & SB(rs)cd              & UGC 12; excellent case   \\                                                                                                                                                
CIG0004         &   279 &    5.0 &  3  & SA(rs)c  sp          & NGC 7817; excellent case;   \\                                                                                                                                              
\phantom{NGC  }" & "\phantom{00} & "\phantom{0} & "\phantom{0} & \phantom{00}" & highly-inclined but not edge-on; like  \\                                                                                                                  
\phantom{NGC  }" & "\phantom{00} & "\phantom{0} & "\phantom{0} & \phantom{00}" & N0253; large (rs)  \\                                                                                                                                      
CIG0005         &   602 &    3.0 & ..  & SA(r$\underline{\rm s}$)b:             & CGCG 456-32; poorly resolved   \\                                                                                                                         
CIG0006         &   652 &    9.0 &  4  & SA$\underline{\rm B}$(s)m      pec     & NGC 9; resembles an RG, but no   \\                                                                                                                       
\phantom{NGC  }" & "\phantom{00} & "\phantom{0} & "\phantom{0} & \phantom{00}" & companion; $m$=1 spiral  \\                                                                                                                                
CIG0007         &   793 &    2.5 &  9  & SAB(rs)a$\underline{\rm b}$              & CGCG 382-30; excellent face-on   \\                                                                                                                     
\phantom{NGC  }" & "\phantom{00} & "\phantom{0} & "\phantom{0} & \phantom{00}" & case; $m$=2 mainly  \\                                                                                                                                     
CIG0008         &   833 &    6.0 &  9  & SA(s)cd              & UGC 111; complex but regular   \\                                                                                                                                           
\phantom{NGC  }" & "\phantom{00} & "\phantom{0} & "\phantom{0} & \phantom{00}" & spiral  \\                                                                                                                                                 
CIG0009         &   859 &    6.0 &  1: & SA(s:)cd:     pec     & UGC 116; large blue associations   \\                                                                                                                                      
CIG0011         &   963 &    6.5 &  5  & SA(s)c$\underline{\rm d}$              & UGC 139; excellent, large   \\                                                                                                                            
\phantom{NGC  }" & "\phantom{00} & "\phantom{0} & "\phantom{0} & \phantom{00}" & late-type spiral  \\                                                                                                                                       
CIG0012         &  1056 &    6.0 & ..  & SAB:(s:)cd: sp  pec     & UGC 149; pointy-ended; warping?   \\                                                                                                                                     
 \hline
 \end{tabular}
 \end{table*}

Table~\ref{tab:comp12} shows that in general, the phase 1 and 2
classifications are very similar. Of the 719 galaxies, 112 (15.5\%)
received identical full classifications in the two phases, very similar
to the 16\% found by Buta et al. (2015). For stage, family, and
variety, more than 50\% of the subsets of the galaxies for which these
aspects could be evaluated received identical classifications. However,
for the outer variety, a significant fraction (36.8\%) are in the
category ``$OV$ 1 no $OV$ 2, or $OV$ 2 no $OV$ 1," which includes all
cases where an outer feature was recognized in one phase, but not in
the other. It appears that more outer features were noticed in the
first phase as compared with the second.  Any recognition of an outer
feature was included in the final adopted average classification.

\begin{table}
\centering
\caption{Comparison of Phase 1 and 2 classifications.  Col. 1: ``Type" = full letter classification [stage ($T$), family ($F$), inner variety ($IV$), and outer variety ($OV$)]. The categories ``$IV$ 1 no $IV$ 2, or $IV$ 2 no $IV$ 1" and ``$OV$ 1 no $OV$ 2, or $OV$ 2 no $OV$ 1" refer to those galaxies where a feature was recognized in one phase, but not in the other.  Col. 2: numbers of galaxies in the categories listed in col. 1. Col. 3: percentages of these numbers out of the total numbers in the $T$, $F$, $IV$, and $OV$ subsamples.}
\label{tab:comp12}
\begin{tabular}{lrr}
\hline
Comparison & $n$& \% of $N$ \\
\hline
Type 1 = Type 2    & 112 & 15.5 \\
                   &     &      \\
$\Delta T$=0       & 429 & 60.5 \\
$|\Delta T|$=1     & 224 & 31.6 \\
$|\Delta T|$=2     &  35 &  4.9 \\
$|\Delta T| >$2    &  21 &  3.0 \\
$N$                & 709 &      \\
                   &     &      \\
$\Delta F$=0.00    & 448 & 73.4 \\
$\Delta F$=0.25    & 116 & 19.0 \\
$\Delta F$=0.50    &  43 &  7.1 \\
$\Delta F$=0.75    &   1 &  0.2 \\
$\Delta F$=1.00    &   2 &  0.3 \\
$N$                & 610 &      \\
                   &     &      \\
$IV$ 1 = $IV$ 2        & 311 & 51.1 \\
$IV$ 1 $\neq$ $IV$ 2   & 249 & 40.9 \\
$IV$ 1 no $IV$ 2, or $IV$ 2 no $IV$ 1 &  49 & 8.0 \\
$N$                & 609 &      \\
                   &     &     \\
$OV$ 1 = $OV$ 2        & 102 & 44.2 \\
$OV$ 1 $\neq$ $OV$ 2   &  44 & 19.0 \\
$OV$ 1 no $OV$ 2, or $OV$ 2 no $OV$ 1 &  85 & 36.8 \\
$N$                & 231 &      \\
                   &     &      \\
\hline
\end{tabular}
\end{table}

The root mean square (rms) dispersion of classifications between
phases 1 and 2 is calculated as

$$\sigma_{p1p2}(T) = {1\over N}\Sigma(T_{p1}-T_{p2})^2   \eqno{1a}$$
$$\sigma_{p1p2}(F) = {1\over N}\Sigma(F_{p1}-F_{p2})^2   \eqno{1b}$$
$$\sigma_{p1p2}(IV) = {1\over N}\Sigma(IV_{p1}-IV_{p2})^2   \eqno{1c}$$

\noindent
where $N$ is the total number of galaxies in each comparison. From the
comparisons shown in Figure~\ref{fig:p12}, and using the numerical
codes from Buta et al. (2015), we obtain $\sigma_{p1p2}(T)$ = 0.89
stage intervals, corresponding to $\sigma_{p1}(T) \approx
\sigma_{p2}(T)$ = 0.63 stage intervals, based on 707 galaxies. Here 1
stage interval means a difference like Sbc to Sc, or S0/a to Sa, so the
internal consistency is good to better than 1 stage interval. For
family, we find $\sigma_{p1p2}(F)$ = 0.72 family intervals, where 1
family interval equals a difference like SA to S$\underline{\rm A}$B or
SA$\underline{\rm B}$ to SB. This corresponds to $\sigma_{p1}(F)
\approx \sigma_{p2}(F)$ = 0.51 family intervals, based on 608
galaxies.

For inner varieties, we obtain $\sigma_{p1p2}(IV)$ = 1.16 variety
intervals, where 1 variety interval equals a difference like (r) to
($\underline{\rm r}$s) or (r$\underline{\rm s}$) to (s). This
corresponds to $\sigma_{p1}(IV) \approx \sigma_{p2}(IV)$ = 0.82 variety
intervals, based on 547 galaxies. The consistency is somewhat poorer
for inner variety because of the many additional categories added by
the recognition of inner ring-lenses (rl) and inner lenses (l), which
are combined in Figure~\ref{fig:p12}.

\section{External Comparison of Classifications}

Galaxy classification at the present time is often done by
consensus and does not always involve estimation of standard types in
the fashion Sa, SBb, SAB(rs)ab, etc.  For example, Fukugita et al.
(2007) classified 2253 SDSS galaxies in a modified Hubble system, based
on the independent examination of all of the sample galaxies by three
astronomers. In this study, only stages and families were judged;
inner, outer, and nuclear varieties were not. Ann, Seo, and Ha (2015)
estimated stages and families of 5836 galaxies having $z$ $<$ 0.01.
Baillard et al. (2011) present classifications for 4458 galaxies in the
EFIGI sample, based on the participation of 10 professional astronomers
who each classified a 10\% subset of the sample, including an overlap
sample to examine and remove personal equations and potential biases.
In this case, the classifications were carried out using numerical
codings of 16 ``attributes," or morphological characteristics.

Kartaltepe et al.  (2016) used a similar procedure to classify 7634
galaxies that are part of the CANDELS survey, a dataset that includes
deep images of galaxies in the redshift range 0 $<$ $z$ $<$ 4. In this
case, 65 astronomers contributed to the final classifications, which
were tailored for the higher redshift part of the sample (i.e., did not
involve classifications like SB(rs)c, SA(s)a, etc.) 

\begin{figure}
\includegraphics[width=\columnwidth]{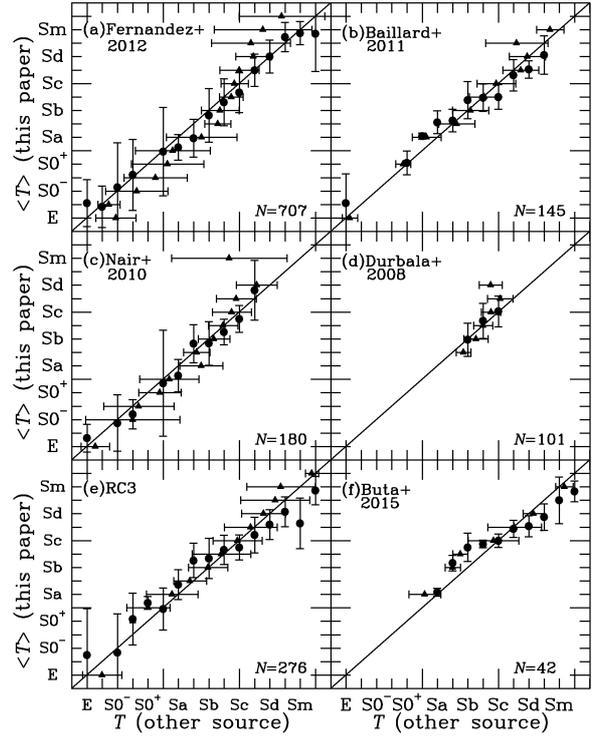}
\caption{Comparison of Table~\ref{tab:catalog} mean Phase 1 and 2 stages 
with similar data from six other sources: (a) Fern\'andez-Lorenzo et al.
(2012); (b) Baillard et al. (2011); (c) Nair \& Abraham (2010);
(d) Durbala et al. (2008); (e) RC3 (de Vaucouleurs et al. 1991);
and (f) Buta et al. (2015).}
\label{fig:newcomps}
\end{figure}

\begin{figure}
\includegraphics[width=\columnwidth]{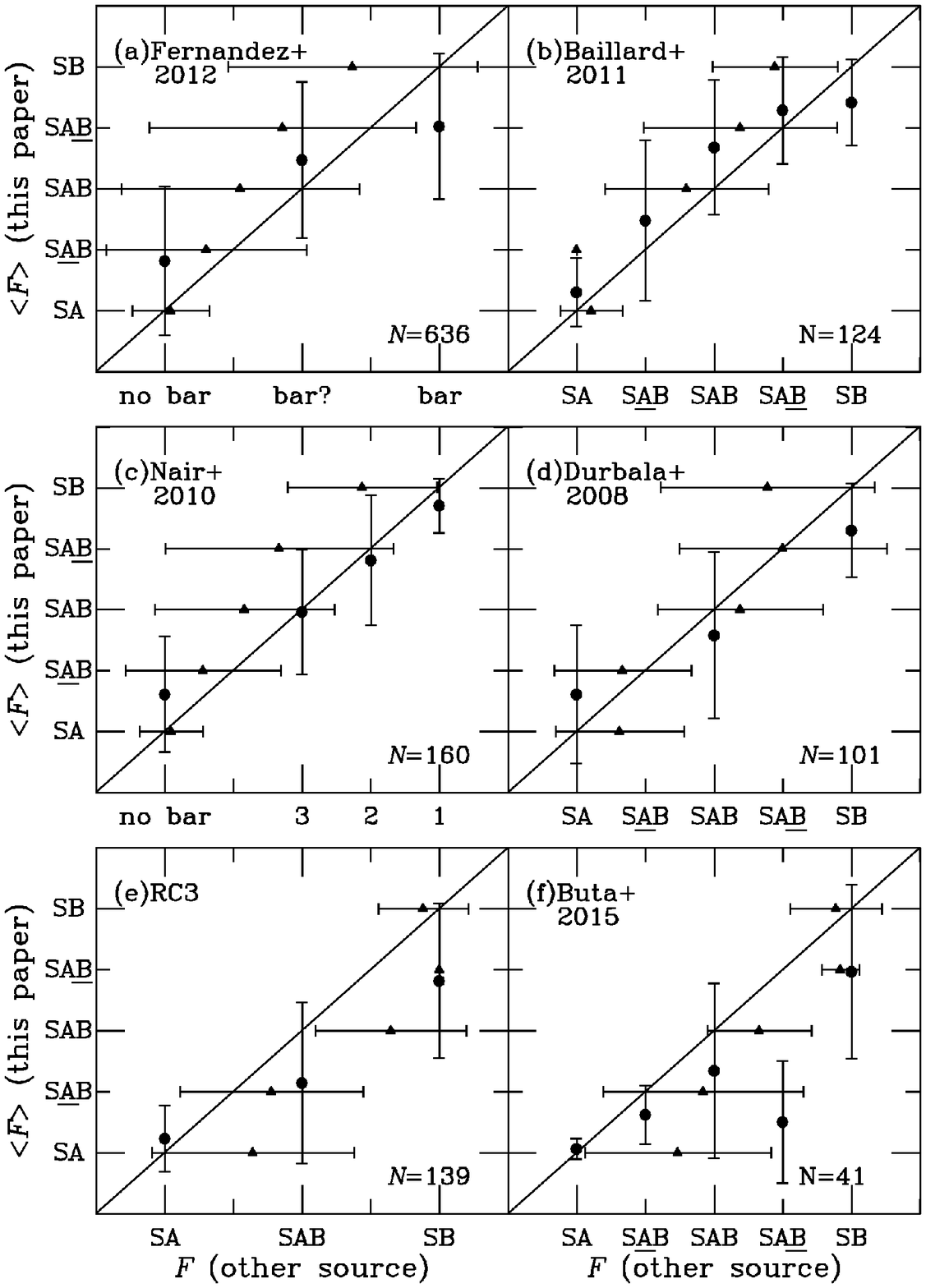}
\caption{Comparison of Table~\ref{tab:catalog} mean Phase 1 and 2 families 
with similar data from six other sources: (a) Fern\'andez-Lorenzo et al.
2012; (b) Baillard et al. (2011); (c) Nair \& Abraham (2010);
(d) Durbala et al. (2008); (e) RC3 (de Vaucouleurs et al. 1991);
and (f) Buta et al. (2015).}
\label{fig:families}
\end{figure}

We have not used the multi-classifier approach in our examination of
the AMIGA sample. To evaluate the external consistency of our
classifications, we compare our $T$ and $F$ estimates with those from
five other sources:  Baillard et al. (2011); the Nair \& Abraham (2010)
sample (14034 SDSS galaxies); Fern\`andez-Lorenzo et al. (2012; all CIG
galaxies); Durbala et al.  (2008; 101 AMIGA galaxies from SDSS); and
RC3.  Figures~\ref{fig:newcomps} and ~\ref{fig:families} show
comparisons between our Table~\ref{tab:catalog} mean phase 1 and 2
classifications and the galaxies in common with these other sources for
stage and family, respectively. These show no serious scale differences
between sources, but the large amount of scatter is consistent with
previous findings. Similar to equations 1, we estimate the root mean
square (rms) dispersion between observers $i$ and $j$ as

$$\sigma_{ij}^2 = {1\over N}\Sigma(T_i-T_j)^2   \eqno{2}$$

\noindent
where again $N$ is the total number of galaxies in the comparison.
If we assume no systematic effects between different observers, then these
rms dispersions are related to the individual dispersions through equations
like 

$$\sigma_{ij}^2 = \sigma_i^2+ \sigma_j^2   \eqno{3}$$

\noindent
This will give 15 equations in 6 unknowns which we solve using linear
least squares. 

%




Table~\ref{tab:external} summarizes the results of these comparisons.
The $\sigma_i$ on stages range from 0.6 to 1.3 stage intervals and can
only be considered approximate because each $\sigma_{ij}$ combination
involves a different subset of the sample, with the number of galaxies
ranging from $N$ = 39 to $N$ = 705. The same is also true for family
classifications. The average external consistency on stages is
$<\sigma(T)>$ $\approx$ 1.1 stage intervals, while the external
consistency on family classifications is 0.24 or $\approx$1 family
interval.

\begin{table*}
\centering
\caption{
External Agreement Between Classifications.  Each column gives the rms dispersion between the two sources ($i$ and $j$) indicated. The number in parentheses next to each value is the number of galaxies in each comparison. The individual $\sigma_i$ are derived from a linear least squares analysis.  Sources: EFIGI=Baillard et al. (2011); NA2010=Nair \& Abraham (2010); Fern2012=Fern\'andez-Lorenzo et al. (2012); Durb2008=Durbala et al. (2008); RC3=de Vaucouleurs et al. (1991).}
\label{tab:external}
\begin{tabular}{cccccccc}
\hline
 Source & $j$ & RB & EFIGI & NA2010 & Sul2006 & Durb2008 & RC3 \\
\hline
      &             &             &             &             &           &   \\
      &             &             &            &  (a) $\sigma_{ij}$ Stage           &           &   \\
      &             &             &             &             &           &   \\
$i$          &  & 1 & 2 & 3 & 4 & 5  & 6 \\
EFIGI &    2 &   1.27(145) &  .......... &  .......... &  .......... &  .......... & .......... \\
NA2010 &    3 &   1.61(180) &   1.44( 57) &  .......... &  .......... &  .......... & .......... \\
Fern2012 &    4 &   1.83(705) &   1.38(147) &   1.89(182) &  .......... &  .......... & .......... \\
Durb2008 &    5 &   1.21(101) &   1.31( 49) &   1.52( 39) &   1.10(101) &  .......... & .......... \\
RC3 &    6 &   1.62(276) &   1.84(147) &   1.57( 60) &   1.83(280) &   1.39( 55) & .......... \\
      &             &             &             &             &           &   \\
$\sigma_i(T)$  & & 1.06 & 0.95 & 1.21 & 1.25 & 0.62 & 1.29 \\
      &             &             &             &             &           &   \\
      &             &             &             & (b) $\sigma_{ij}$ Family            &           &   \\
      &             &             &             &             &           &   \\
EFIGI &    2 &   0.24(124) &  .......... &  .......... &  .......... &  .......... & .......... \\
NA2010 &    3 &   0.29(161) &   0.20( 54) &  .......... &  .......... &  .......... & .......... \\
Sul2006 &    4 &   0.36(637) &   0.29(131) &   0.31(176) &  .......... &  .......... & .......... \\
Durb2008 &    5 &   0.32(101) &   0.30( 49) &   0.35( 39) &   0.38(101) &  .......... & .......... \\
RC3 &    6 &   0.39(149) &   0.43( 82) &   0.38( 32) &   0.54(149) &   0.44( 48) & .......... \\
      &             &             &             &             &           &   \\
$\sigma_i(F)$  & & 0.19 & 0.14 & 0.16 & 0.30 & 0.25 & 0.38 \\
\hline
\end{tabular}
\end{table*}

Naim et al. (1995) carried out an experiment to examine the external
consistency in morphological classifications between different
observers.  Using paper copies of blue-light images (or monitor
displays) of 831 galaxies, six observers classified the galaxies in
modified Hubble systems.  Although general consistency in stage
classifications between observers was found, a non-negligible scatter
was also found with an average $\sigma_{ij}$ $\approx$ 1.8 stage
intervals. Our analysis in Table~\ref{tab:external} has a
$<\sigma_{ij}>$ =1.52, which may be a little better because of improved
image quality.

In general, $\sigma(T)$ = 1.1 can be considered ``good" for Hubble
classifications from different sources. It means that a galaxy
classified as type Sbc from one source could be classified as Sb or Sc
by another source. This level of disagreement is relatively small
compared to the 16-stage extent of the VRHS sequence.

\section{Morphological Characteristics of the Sample}

\subsection{Stages}

\begin{figure}
\includegraphics[width=\columnwidth]{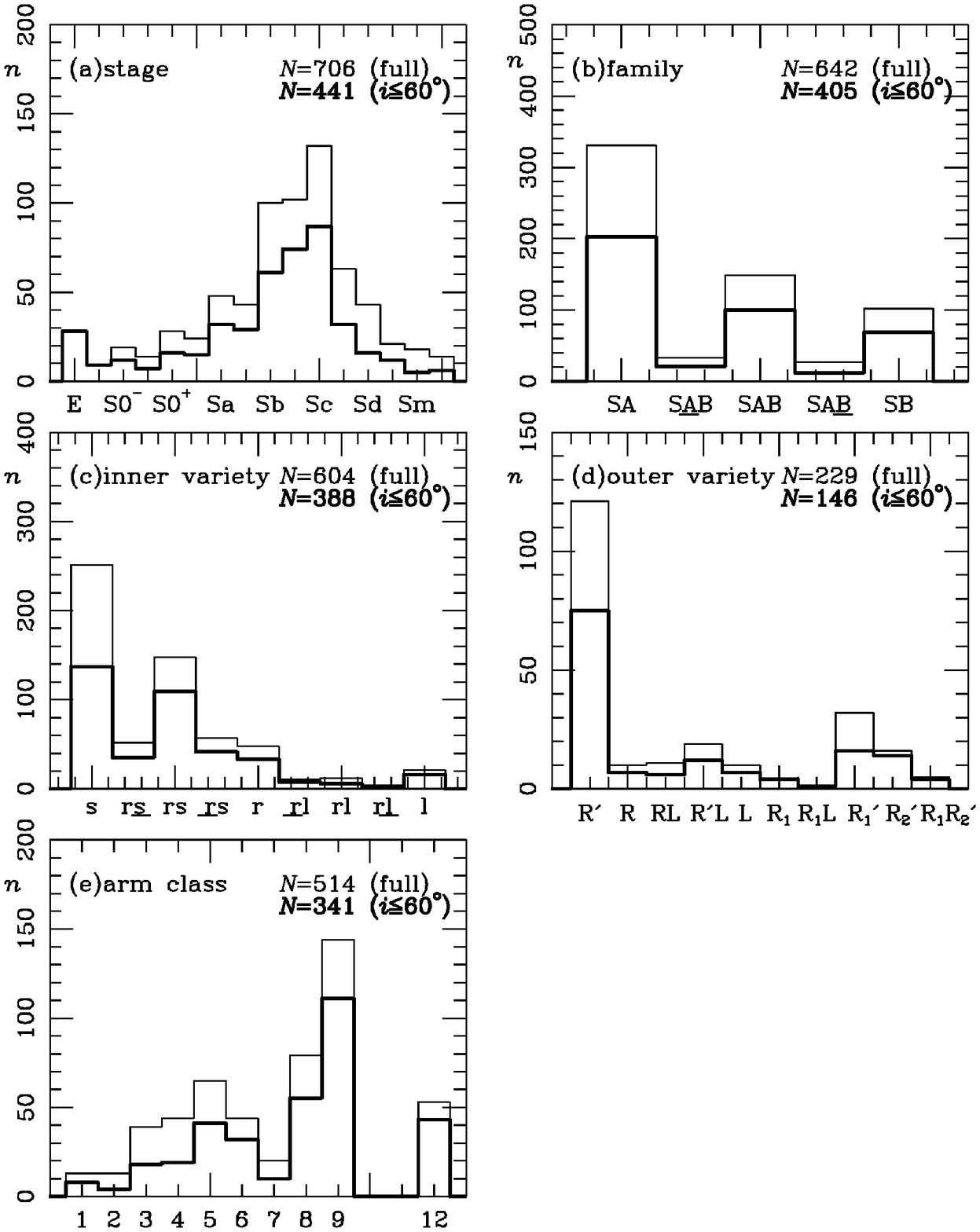}
\caption{Histograms of morphological classifications from Table~\ref{tab:histdata}
for the full sample (light solid histogram) and a subsample restricted to inclinations
$i$ less than or equal to 60$^o$ (heavy solid histogram). The numbers of objects $N$ in each sample
are indicated.}
\label{fig:histos}
\end{figure}

Table~\ref{tab:histdata} lists our revised classification results,
which are plotted as histograms in Figure~\ref{fig:histos}. The
distributions of stages, families, inner varieties, outer varieties and
Elmegreen arm classes are compiled for two samples: the full set of
AMIGA galaxies classified, and a subset restricted to more face-on
disks (inclination $i$ $\leq$ 60$^o$). The latter restriction is important
because features such as bars and rings can become harder to recognize 
when the inclination is high. 

The distribution of stages confirms the finding of Sulentic et al.
(2006; see also Fern\'andez-Lorenzo et al. 2012) that the most abundant
types among isolated galaxies are Sb-Sc spirals. However, while these
constitute 63\% of the Sulentic et al. (2006) sample and 68\% of the
Fern\'andez-Lorenzo et al. (2012) sample, they make up 47\%-50\% of our
current sample. We also find Sa-Sd spirals at the 75\% level, compared
to 82\% for Sulentic et al. (2006). For E and S0 galaxies, we find
13.9\% for the full sample and 16.3\% for the inclination-restricted
subset, both comparable to what was found by Sulentic et al. (2006).
The higher value for the restricted subset is partly or perhaps wholly
due to the rejection of spindle galaxies.  In general, the results from
the restricted subset are very similar to those from the full sample.

Sulentic et al. (2006) noted a low fraction of Sa galaxies in their
sample.  Including S0/a types, their classification has 3.2\%
early-type spirals compared to 10.3\%-10.7\% in our classification. In
their sample also, Sdm-Im types make up 5.6\% of the galaxies, while
these make up 7.5\%-2.5\% of our sample. Considering the full range of
types from S0/a to Sm, spirals represent 84.2\%-82.3\% of our AMIGA
sample.

Of the 706 galaxies in the sample for which a stage could be judged,
18\% have a CVRHS classification in Table~\ref{tab:catalog} with the
appendage ``pec," implying something ``peculiar" or unusual, most
likely an asymmetry. For some objects, the classification is ``Pec
(merger)", meaning the object could be a merger of two galaxies.
Although it is tempting to conclude that such cases must therefore not
be truly isolated, the presence of peculiarities is not an automatic
disqualifier from our catalogue. This is because isolation only depends
on neighborhood, and we can ask if some peculiarities could arise in
isolation. Objects classified as ``Pec (merger)" are by default not
included in Table~\ref{tab:histdata} since they have no stage, family,
inner variety, or outer variety as part of their classification.

Eleven sample objects (CIG 31, 424, 468, 532, 533, 678, 773, 893, 927,
and 1038) have apparent close companions. Without redshift information
for many of the galaxies, and without perturbations pointing straight
to the potential companion, we cannot affirm that most of these cases
involve real companions. An exception is CIG 533, for which redshift
information is available for both the galaxy and the companion.

\begin{table*}
\centering
\caption{Histogram Data for Morphological Analysis. The full sample includes all the galaxies in the AMIGA sample that were classifiable into the given categories. The restricted subset excludes galaxies having an inclination $i$
$>$ 60$^o$.}
\label{tab:histdata}
\begin{tabular}{lrlrlrlr}
\hline
Stage & \% ($n$) & Family  & \% ($n$) & Outer   & \% ($n$) & Arm   & \% ($n$) \\ 
      &          & Variety &          & Feature &          & Class &      \\
\hline
                          &             &                           &             &                           &             & &      \\
\noalign{\centerline{Full sample}}
                          &             &                           &             &                           &             & &      \\
E                         &   4.0 ( 28) & SA                        &  51.6 (331) & (R$^{\prime}$)            &  52.8 (121) & AC  1                     &   2.5 ( 13) \\
E$^+$                     &   1.3 (  9) & S$\underline{\rm A}$B     &   5.1 ( 33) & (R)                       &   4.4 ( 10) & AC  2                     &   2.5 ( 13) \\
S0$^-$                    &   2.7 ( 19) & SAB                       &  23.2 (149) & (RL)                      &   4.8 ( 11) & AC  3                     &   7.6 ( 39) \\
S0$^o$                    &   2.0 ( 14) & SA$\underline{\rm B}$     &   4.2 ( 27) & (R$^{\prime}$L)           &   8.3 ( 19) & AC  4                     &   8.6 ( 44) \\
S0$^+$                    &   4.0 ( 28) & SB                        &  15.9 (102) & (L)                       &   4.4 ( 10) & AC  5                     &  12.6 ( 65) \\
S0/a                      &   3.4 ( 24) & S\uAB+SAB+S\AuB+SB     &  48.4 (311) & (R$_1$)                   &   1.7 (  4) & AC  6                     &   8.6 ( 44) \\
Sa                        &   6.8 ( 48) & $N$                       & 642 & (R$_1$L)                  &   0.4 (  1) & AC  7                     &   3.9 ( 20) \\
Sab                       &   6.1 ( 43) & .......                   &....... & (R$_1^{\prime}$)          &  14.0 ( 32) & AC  8                     &  15.4 ( 79) \\
Sb                        &  14.2 (100) & (s)                       &  41.6 (251) & (R$_2^{\prime}$)          &   7.0 ( 16) & AC  9                     &  28.0 (144) \\
Sbc                       &  14.4 (102) & (r$\underline{\rm s}$)    &   8.6 ( 52) & (R$_1$R$_2^{\prime}$)     &   2.2 (  5) & AC 12                     &  10.3 ( 53) \\
Sc                        &  18.7 (132) & (rs)                      &  24.5 (148) & R$_1$+R$_2$               &  25.3 ( 58) & AC1-AC4                   &  21.2 (109) \\
Scd                       &   8.9 ( 63) & ($\underline{\rm r}$s)+(r$^{\prime}$l)    &   9.4 ( 57) & $N$                       & 229 & AC5-AC8                   &  40.5 (208) \\
Sd                        &   6.1 ( 43) & (r)                       &   7.9 ( 48) & .......                   &....... & AC9,AC12                  &  38.3 (197) \\
Sdm                       &   3.0 ( 21) & ($\underline{\rm r}$l)    &   1.8 ( 11) & .......                   &....... & $N$                       & 514 \\
Sm                        &   2.5 ( 18) & (rl)                      &   2.0 ( 12) & .......                   &....... & .......                   &....... \\
Im                        &   2.0 ( 14) & (r$\underline{\rm l}$)    &   0.7 (  4) & .......                   &....... & .......                   &....... \\
E-S0$^+$                  &  13.9 ( 98) & (l)                       &   3.5 ( 21) & .......                   &....... & .......                   &....... \\
Sa-Sd                     &  75.2 (531) & \rus+(rs)+\urs+(r$^{\prime}$l)+(r)        &  50.5 (305) & .......                   &....... & .......                   &....... \\
Sb-Sc                     &  47.3 (334) & \uurl+(rl)+\rul+(l)        &   7.9 ( 48) & .......                   &....... & .......                   &....... \\
$N$                       & 706 & $N$                       & 604 & .......                   &....... & .......                   &....... \\
                          &             &                           &             &                           &             & &      \\
\noalign{\centerline{Restricted to $i\leq$60$^o$}}
                          &             &                           &             &                           &             & &      \\
E                         &   6.3 ( 28) & SA                        &  50.1 (203) & (R$^{\prime}$)            &  51.4 ( 75) & AC  1                     &   2.3 (  8) \\
E$^+$                     &   2.0 (  9) & S$\underline{\rm A}$B     &   5.2 ( 21) & (R)                       &   4.8 (  7) & AC  2                     &   1.2 (  4) \\
S0$^-$                    &   2.7 ( 12) & SAB                       &  24.7 (100) & (RL)                      &   4.1 (  6) & AC  3                     &   5.3 ( 18) \\
S0$^o$                    &   1.6 (  7) & SA$\underline{\rm B}$     &   3.0 ( 12) & (R$^{\prime}$L)           &   8.2 ( 12) & AC  4                     &   5.6 ( 19) \\
S0$^+$                    &   3.6 ( 16) & SB                        &  17.0 ( 69) & (L)                       &   4.8 (  7) & AC  5                     &  12.0 ( 41) \\
S0/a                      &   3.4 ( 15) & S\uAB+SAB+S\AuB+SB     &  49.9 (202) & (R$_1$)                   &   2.7 (  4) & AC  6                     &   9.4 ( 32) \\
Sa                        &   7.3 ( 32) & $N$                       & 405 & (R$_1$L)                  &   0.7 (  1) & AC  7                     &   2.9 ( 10) \\
Sab                       &   6.6 ( 29) & .......                   &....... & (R$_1^{\prime}$)          &  11.0 ( 16) & AC  8                     &  16.1 ( 55) \\
Sb                        &  13.8 ( 61) & (s)                       &  35.3 (137) & (R$_2^{\prime}$)          &   9.6 ( 14) & AC  9                     &  32.6 (111) \\
Sbc                       &  16.8 ( 74) & (r$\underline{\rm s}$)    &   9.0 ( 35) & (R$_1$R$_2^{\prime}$)     &   2.7 (  4) & AC 12                     &  12.6 ( 43) \\
Sc                        &  19.7 ( 87) & (rs)                      &  28.1 (109) & R$_1$+R$_2$               &  26.7 ( 39) & AC1-AC4                   &  14.4 ( 49) \\
Scd                       &   7.3 ( 32) & ($\underline{\rm r}$s)+(r$^{\prime}$l)    &  10.8 ( 42) & $N$                       & 146 & AC5-AC8                   &  40.5 (138) \\
Sd                        &   3.6 ( 16) & (r)                       &   8.5 ( 33) & .......                   &....... & AC9,AC12                  &  45.2 (154) \\
Sdm                       &   2.7 ( 12) & ($\underline{\rm r}$l)    &   2.1 (  8) & .......                   &....... & $N$                       & 341 \\
Sm                        &   1.1 (  5) & (rl)                      &   1.5 (  6) & .......                   &....... & .......                   &....... \\
Im                        &   1.4 (  6) & (r$\underline{\rm l}$)    &   0.5 (  2) & .......                   &....... & .......                   &....... \\
E-S0$^+$                  &  16.3 ( 72) & (l)                       &   4.1 ( 16) & .......                   &....... & .......                   &....... \\
Sa-Sd                     &  75.1 (331) & \rus+(rs)+\urs+(r$^{\prime}$l)+(r)        &  56.4 (219) & .......                   &....... & .......                   &....... \\
Sb-Sc                     &  50.3 (222) & \uurl+(rl)+\rul+(l)        &   8.2 ( 32) & .......                   &....... & .......                   &....... \\
$N$                       & 441 & $N$                       & 388 & .......                   &....... & .......                   &....... \\
\hline
\end{tabular}
\end{table*}

The distribution of stages for the AMIGA sample is very different from that 
for the S$^4$G sample. Figure 5 of Buta et al. (2015) shows that the latter
sample strongly emphasizes extreme late-type disk galaxies, i.e., galaxies
in the stage range Sd-Im. These constitute 48.5\%$\pm$1.4\% of 1240 low
inclination galaxies in the S$^4$G sample, compared to 16.1\%$\pm$1.8\%
for the restricted subset of AMIGA galaxies.

\subsection{Bar Classifications and Fraction}

\begin{figure}
\includegraphics[width=\columnwidth]{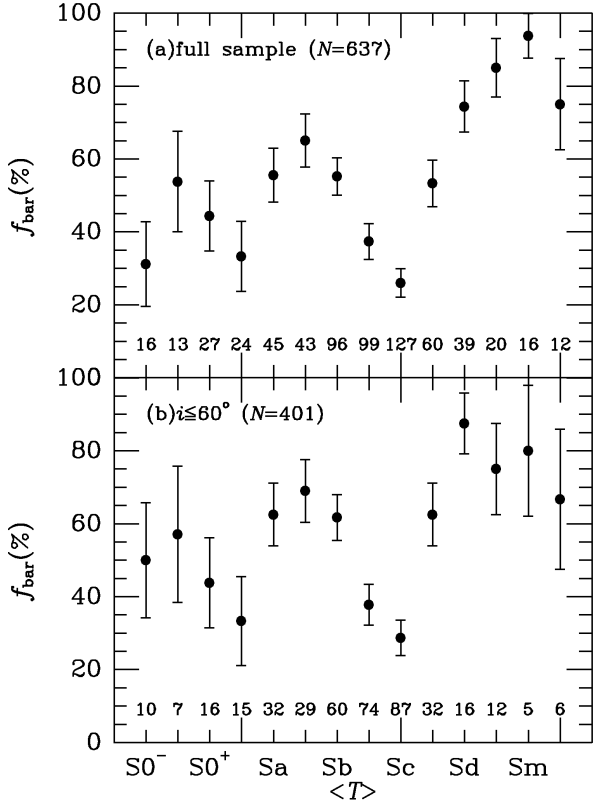}
\caption{Graphs of the bar fraction $f_{bar}$ versus the mean phase 1
and 2 stage along the CVRHS Hubble sequence: (a) for the full sample
unrestricted by inclination, and (b) for the restricted subset of
galaxies having $i$ less than or equal to 60$^o$. The number of
objects $n$ at each stage is indicated.}
\label{fig:pbar}
\end{figure}

The bar fraction for isolated galaxies is clearly of interest. Verley
et al. (2007c) observed a sample of 45 well-resolved, low inclination
CIG galaxies and found that 60\% are barred and 33\% are nonbarred.
Here we define the bar fraction in two ways:  (1) $f_{1bar}$ =
$100*[N-n({\rm SA})-n({\rm S}\underline{\rm A}{\rm B})]/N$, where $N$
is the total number of galaxies classifiable as to family, $n$({\rm
SA}) = number of SA galaxies, and $n$({\rm S\uAB}) = number of
S\uAB\ galaxies. This is how the bar fraction was defined by Buta et
al. (2015). (2) $f_{2bar}$ = $100*[N-n(SA)]/N$ - This definiton allows
for a fairer comparison with $f_{bar}$ in the mid-IR. Bars tend to
generally look stronger in IR light (the ``stronger bar effect"), and
this definition assumes that an S\uAB\ galaxy in the $g$-band might be
classified as SAB in the mid-IR. This allows the bar fraction to
include both the strongest and the weakest-looking bars. With these
definitions and the data in Table~\ref{tab:histdata}, the bar fraction
is $f_{1bar}$ = 45\%$\pm$3\% and $f_{2bar}$=50.0\%$\pm$3\%, both for
the restricted (low inclination) subset. These values are lower than
the 62\%-71\% found by Buta et al.  (2015) for the S$^4$G mid-infrared
sample, and for the Ohio State University Bright Galaxy Survey
(Eskridge et al. 2000).  These percentages are over all types.
As shown in Figure 7 of Buta et al. (2015), the bar fraction has a
minimum of $\approx$40\% in the stage range S$\underline{\rm b}$c-Sc,
the same range where the bulk of isolated spirals are found. The
equivalent version of Figure 7 of Buta et al. (2015) for the full AMIGA
sample and its inclination-restricted subset is shown in
Figures~\ref{fig:pbar}a and b, respectively. As for the S$^4$G sample,
our AMIGA sample shows a minimum bar fraction in the stage range
Sbc-Sc, ranging from 37\%$\pm$5\% at stage Sbc to 26\%$\pm$4\% at stage
Sc. These numbers change to 38\%$\pm$6\% and 29\%$\pm$5\% for the
restricted subset. Only 16\% of our AMIGA galaxies were classified as
family ``SB" in the full sample and 17\% in the restricted subset.

Table~\ref{tab:ivars} summarizes the bar fraction of AMIGA galaxies
over the same ranges of type as in Table 7 of Buta et al. (2015). If we
compare the restricted AMIGA sample to the equivalent S$^4$G subset
(infrared axis ratio $q_{25.5}$$\leq$0.5), for stages S0/a to Sc, the
result is $f_{1bar}$ = 40\%$\pm$3\% for AMIGA galaxies versus
55\%$\pm$2\% for S$^4$G galaxies. For stages Scd to Sm, the fractions
are both higher: $f_{1bar}$ = 66\%$\pm$6\% for AMIGA versus
81\%$\pm$2\% for S$^4$G. The bar fraction of AMIGA galaxies appears to
be significantly lower than in S$^4$G galaxies; however, especially for
the S0/a-Sc stage range, the difference could be partly attributable to
the ``stronger bar" effect in mid-IR images. Thus, $f_{2bar}$ is
probably a better definition for $g$-band classifications.
Table~\ref{tab:ivars} shows that $f_{2bar}$ = 46\%$\pm$3\% for S0/a to
Sc and 72\%$\pm$6\% for Scd-Sm. Even allowing for the different
definitions of $f_{bar}$, the AMIGA sample has a slightly lower bar
fraction than does the S$^4$G sample.

Over the full type range of S0/a to Sm, a more significant difference
emerges: $f_{2bar}$ = 50\%$\pm$3\% for AMIGA galaxies versus $f_{1bar}$
= 66\%$\pm$2\% for S$^4$G galaxies. Much of this difference is
attributable to the emphasis of the AMIGA sample on Sb-Sc galaxies, as
compared to Scd-Im galaxies in S$^4$G. The latter range pulls up the
bar fraction in the mid-IR substantially.

Figures~\ref{fig:p12}--~\ref{fig:families}  show that, while our Phase
1 and 2 bar classifications are internally consistent, systematic
disagreements with other sources of bar classifications are present. In
comparison with RC3 classifications, there is a trend for weaker bar
classifications in Table~\ref{tab:catalog}. Compared to Baillard et al.
(2011) and Nair \& Abraham (2010), however, there is a slight trend for
stronger bar classifications in Table~\ref{tab:catalog}.
Fern\'andez-Lorenzo et al. (2012) were able to classify bars only as
being ``bar", ``bar?", and ``no bar",  which we have treated as SB,
SAB, and SA, respectively.  For these we find reasonably good agreement
(Figure~\ref{fig:families}a).

The graphs in Figures~\ref{fig:newcomps}f and ~\ref{fig:families}f
show the comparisons between Table~\ref{tab:catalog} types and those
based on mid-IR images from Buta et al. (2015). Only 42 galaxies are in
common between the S$^4$G sample and our AMIGA sample. The comparisons
show both an ``earlier effect" and a ``stronger bar effect" between the
3.6$\mu$m classifications and our Table~\ref{tab:catalog} $g$-band
classifications, which is not unexpected.

\subsection{Inner and Outer Varieties}

\begin{figure}
\includegraphics[width=\columnwidth]{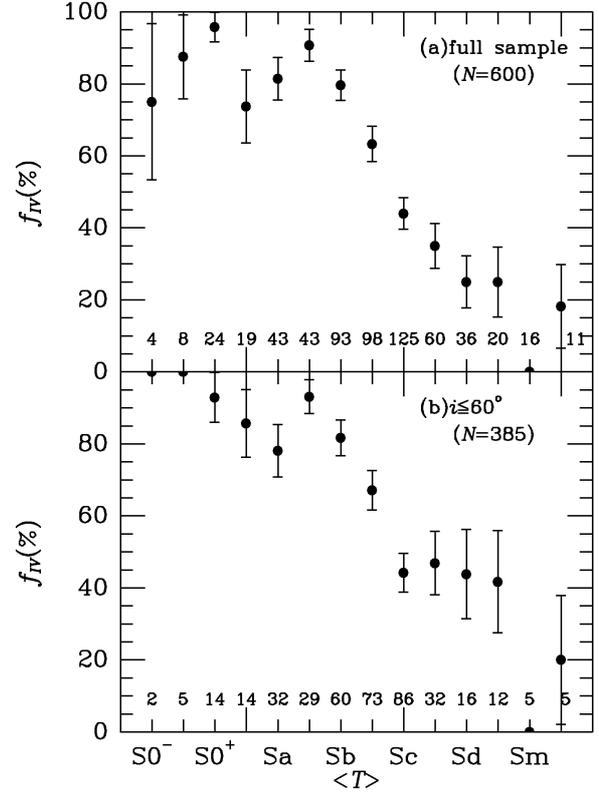}
\caption{Graphs of the inner ring/pseudoring/lens fraction $f_{IV}$ versus the mean phase 1
and 2 stage along the CVRHS Hubble sequence: (a) for the full sample
unrestricted by inclination, and (b) for the restricted subset of
galaxies having $i$ less than or equal to 60$^o$. The number of
objects $n$ at each stage is indicated.}
\label{fig:pvar}
\end{figure}

The results for inner variety ($IV$) classifications in
Table~\ref{tab:histdata} show that the dominant variety is (s), which
is also characteristic of the S$^4$G sample. Figure~\ref{fig:pvar}
shows the relative frequency of the non-(s) varieties versus the mean
phase 1 and 2 stage. As in Figure 8 of Buta et al. (2015), this
frequency is highest among early-type disk galaxies and lowest among
the very latest types. Table~\ref{tab:ivars} summarizes the frequencies
for different ranges of stage and feature types. For the stage range
S0/a to Sc, the restricted AMIGA subset has $f_{IV}$ = 54\%$\pm$3\% for
(rs)+($\underline{\rm r}$s)+(r$^{\prime}$l)+(r) versus 51\%$\pm$2\% for
the same types for S$^4$G (Table 8 of Buta et al.  2015). For Scd-Sm,
the numbers are $f_{IV}$ = 23\%$\pm$5\% versus 13\%$\pm$2\%,
respectively. Thus, the AMIGA sample has  a slightly higher percentage
of rings and pseudorings compared to S$^4$G. As for the bar fraction,
the difference is more significant when the full type range (S0/a-Sm)
is considered:  49\%$\pm$3\% for AMIGA versus 34\%$\pm$2\$ for S$^4$G.
Again, this is largely due to the strong emphasis of the S$^4$G sample
on later type galaxies as compared to the AMIGA sample.

\begin{table*}
\centering
\caption{Relative frequencies of bars and inner rings in the AMIGA sample, in intervals of stage.
}
\label{tab:ivars}
\begin{tabular}{lrrrrr}
\hline
Parameter & Features & S0$^-$--S0$^+$ & S0/a--Sc & Scd-Sm & S0/a-Sm \\ 
\hline
 $f_{bar}$\% (full)            &  SAB+S\AuB+SB &  37.5$\pm$ 6.5( 21) &  37.1$\pm$ 2.3(161) &  64.4$\pm$ 4.1( 87) &  43.6$\pm$ 2.1(248) \\ 
 $f_{bar}$\% (full)           & S\uAB+SAB+S\AuB+SB &  42.9$\pm$ 6.6( 24) &  42.4$\pm$ 2.4(184) &  68.9$\pm$ 4.0( 93) &  48.7$\pm$ 2.1(277) \\ 
$N$  &     &    56 &   434 &   135 &   569 \\
 $f_{bar}$\% ($i$$\leq$60$^o$) & SAB+S\AuB+SB &  47.1$\pm$ 8.6( 16) &  39.7$\pm$ 2.8(118) &  66.2$\pm$ 5.9( 43) &  44.5$\pm$ 2.6(161) \\ 
 $f_{bar}$\% ($i$$\leq$60$^o$) & S\uAB+SAB+S\AuB+SB &  47.1$\pm$ 8.6( 16) &  45.5$\pm$ 2.9(135) &  72.3$\pm$ 5.6( 47) &  50.3$\pm$ 2.6(182) \\ 
$N$ &      &    34 &   297 &    65 &   362 \\
$f_{IV}$\% (full)            & rs+$\underline{\rm r}$s+r$^{\prime}$l+r & 38.9$\pm$ 8.1( 14) &  51.8$\pm$ 2.4(218) &  14.4$\pm$ 3.1( 19) &  42.9$\pm$ 2.1(237) \\ 
 $f_{IV}$\% (full)            & $\underline{\rm r}$l+rl+r$\underline{\rm l}$+l &  52.8$\pm$ 8.3( 19) &   5.9$\pm$ 1.2( 25) &   0.8$\pm$ 0.8(  1) &   4.7$\pm$ 0.9( 26) \\ 
$N$ &  &  36 &   421 &   132 &   553 \\
$f_{IV}$\% ($i$$\leq$60$^o$) & rs+$\underline{\rm r}$s+r$^{\prime}$l+r & 42.9$\pm$10.8(  9) &  54.1$\pm$ 2.9(159) &  23.1$\pm$ 5.2( 15) &  48.5$\pm$ 2.6(174) \\ 
 $f_{IV}$\% ($i$$\leq$60$^o$) & $\underline{\rm r}$l+rl+r$\underline{\rm l}$+l &  52.4$\pm$10.9( 11) &   5.8$\pm$ 1.4( 17) &   1.5$\pm$ 1.5(  1) &   5.0$\pm$ 1.2( 18) \\ 
$N$ &  &  21 &   294 &    65 &   359 \\
\hline
\end{tabular}
\end{table*}

The most common outer variety classification in Table~\ref{tab:catalog}
is no outer feature recognized. Among those cases where an outer
feature is recognized (229 in the full sample, and 146 in the
restricted subset), the most common type of feature is an outer
pseudoring, R$^{\prime}$. These are made of outer arms whose variable
pitch angle leads to a ring-like pattern. Combined with outer
pseudoring-lenses, R$^{\prime}$L, these features are found in
56\%$\pm$4\% of the restricted AMIGA subset as compared to 53\%$\pm$3\%
for the same features in the S$^4$G low inclination subset.  In
comparison, closed outer rings (R) are rare in our sample. This is
easily explained by the distribution of stages: the emphasis of
our sample on intermediate-to-late-type spirals favours outer
pseudorings over closed outer rings (e.g., see also Buta \& Combes
1996).

Other kinds of outer features are also seen in AMIGA galaxies,
including what Buta \& Crocker (1991) and Buta (1995) called the outer
Lindblad resonance subclasses R$_1$, R$_1^{\prime}$, R$_2^{\prime}$,
and R$_1$R$_2^{\prime}$ (now called ``outer resonant subclasses"; Buta
2017). In the full sample there are 58 cases, while in the restricted
subset there are 39; these correspond to 25\% and 27\%, respectively,
of those samples. In contrast, these features were found in 10\%$\pm$2\%
of 283 S$^4$G galaxies classified as having an outer feature.  The
closed outer ring (R), outer ring-lens (RL), outer pseudoring-lens
(R$^{\prime}$L), and outer lens (L) galaxies constitute 14\%$\pm$2\% of
the full sample and 14\%$\pm$3\% of the restricted subset. Pseudorings
are still the most common features even among these categories:
R$_1^{\prime}$ for the outer resonant classes and R$^{\prime}$L for the
outer ring-lens classes.

\subsection{Arm Classes}

\begin{figure}
\includegraphics[width=\columnwidth]{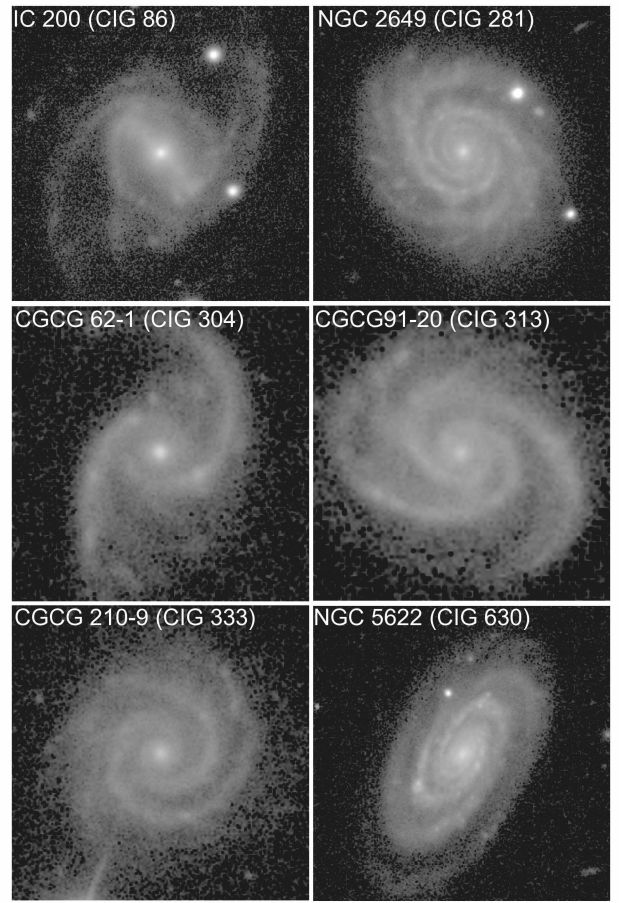}
\caption{Six grand design spirals (CIG 86, 281, 304, 313, 333, 630)
from the AMIGA sample.}
\label{fig:gds}
\end{figure}

The distribution of arm classes shows that grand design spirals (AC8,
AC9, AC12) occur in 53.7\% of the classifiable full sample cases and
61.3\% of the classifiable restricted subset cases. In contrast,
flocculent spirals (AC1-4) occur in 20.2\% of the full sample cases,
and 14.4\% of the restricted subset. Figure~\ref{fig:gds} shows six of
the grand design cases. One case, CIG 86, has a strong bar that could
drive its grand-design pattern (Kormendy and Norman 1979). The other
five, however, are mostly nonbarred, and in fact 50\% of the 276 AC 8,
9, and 12 galaxies in the full sample are nonbarred. While all six of
the galaxies in Figure~\ref{fig:gds} have no significant companions,
four do have small companions (not in the field covered by the
images) ranging from $\approx$1 galaxy diameter away (CIG 304, 313,
630) to more than 5 diameters for CIG 333. CIG 281 has no similar small
companions. Most if not all of the ``knots" seen in these images
are likely star forming regions or foreground stars rather than
companion galaxies. The high abundance of grand design, nonbarred
spirals in the AMIGA sample is an important observation, because it
favours the possibility that such spirals have arisen purely from
internal effects, rather than external interactions.

\begin{table}
\centering
\caption{Morphology and stellar mass in isolated galaxies. Col. 1:
range of specific morphological types or characteristics; col. 2: 
average log stellar mass in units of solar masses; col. 3: standard
deviation of mean log stellar mass; col 4: mean stage of subset;
col. 5: standard deviation of mean stage; col. 6: number of galaxies 
in subset.
}
\label{tab:smasses}
\begin{tabular}{lrrrrr}
\hline
Morphology range & $<log{M_s \over M_{\odot}}>$ & $\sigma_1$ & $<T>$ & $\sigma_1$ & $n$ \\ 
1 & 2 & 3 & 4 & 5 & 6 \\
\hline
    &                              &            &       & &     \\
Full sample      &                              &            &       & &     \\
    &                              &            &       & &     \\
E--S0$^+$                                &   10.72 &    0.36 &   -3.0 &    1.5 &    54\\
S$\underline{\rm 0}$/a--Sab                              &   10.72 &    0.30 &    1.1 &    0.7 &    81\\
Sa$\underline{\rm b}$--Sc                               &   10.58 &    0.32 &    4.0 &    0.8 &   220\\
S$\underline{\rm c}$d--Im                               &    9.80 &    0.55 &    7.1 &    1.4 &    89\\
SA                                       &   10.52 &    0.45 &    3.8 &    7.1 &   209\\
S\uAB,SAB,S\AuB                          &   10.42 &    0.52 &    3.7 &    2.7 &   139\\
SB                                       &   10.31 &    0.65 &    4.2 &    3.2 &    61\\
s,r$\underline{\rm s}$                   &   10.27 &    0.55 &    5.0 &    2.2 &   195\\
rs                                       &   10.57 &    0.38 &    3.6 &    1.6 &    94\\
$\underline{\rm r}$s,r                   &   10.72 &    0.32 &    1.9 &    1.7 &    69\\
$\underline{\rm r}$l,rl,r$\underline{\rm l}$,l  &   10.75 &    0.30 &    0.1 &    2.5 &    28\\
\RP                                   &   10.53 &    0.52 &    3.6 &    2.0 &    73\\
R,RL,R$^{\prime}$L,L                          &   10.67 &    0.44 &    1.0 &    2.6 &    35\\
R$_1$,R$_1$L,R$_1^{\prime}$,R$_2^{\prime}$,R$_1$R$_2^{\prime}$  &   10.74 &    0.18 &    1.8 &    1.2 &    35\\
AC 1--4                                  &   10.13 &    0.59 &    5.7 &    1.9 &    61\\
AC 5--7                                  &   10.40 &    0.40 &    4.6 &    1.4 &    84\\
AC 8,9,12                                &   10.63 &    0.34 &    3.4 &    1.4 &   179\\
    &                              &            &       & &     \\
Restricted to $i$$\leq$60$^o$            &         &         &       & & \\
    &                              &            &       & &     \\
E--S0$^+$                                &   10.75 &    0.33 &   -3.0 &    1.5 &    31\\
S$\underline{\rm 0}$/a--Sab                              &   10.75 &    0.30 &    1.1 &    0.7 &    48\\
Sa$\underline{\rm b}$--Sc                               &   10.60 &    0.31 &    4.0 &    0.8 &   146\\
S$\underline{\rm c}$d--Im                               &    9.84 &    0.50 &    6.9 &    1.4 &    37\\
SA                                       &   10.55 &    0.46 &    3.5 &    2.4 &   120\\
S\uAB,SAB,S\AuB                          &   10.50 &    0.42 &    3.3 &    2.4 &    89\\
SB                                       &   10.50 &    0.46 &    3.4 &    3.0 &    40\\
s,r$\underline{\rm s}$                   &   10.36 &    0.50 &    4.7 &    1.9 &   105\\
rs                                       &   10.55 &    0.38 &    3.5 &    1.6 &    72\\
$\underline{\rm r}$s,r                                  &   10.77 &    0.27 &    2.0 &    1.8 &    46\\
$\underline{\rm r}$l,rl,r$\underline{\rm l}$,l  &   10.73 &    0.33 &    0.3 &    2.7 &    19\\
\RP                                   &   10.56 &    0.47 &    3.3 &    2.0 &    45\\
R,RL,R$^{\prime}$L,L                  &   10.74 &    0.37 &    1.1 &    2.1 &    22\\
R$_1$,R$_1$L,R$_1^{\prime}$,R$_2^{\prime}$,R$_1$R$_2^{\prime}$   &   10.76 &    0.19 &    1.9 &    1.3 &    22\\
AC 1--4                                  &   10.10 &    0.53 &    5.6 &    1.8 &    26\\
AC 5--7                                  &   10.42 &    0.33 &    4.6 &    1.3 &    51\\
AC 8,9,12                                &   10.64 &    0.34 &    3.3 &    1.4 &   137\\
\hline
\end{tabular}
\end{table}

\section{Morphology and Stellar Mass}

\begin{figure}
\includegraphics[width=\columnwidth]{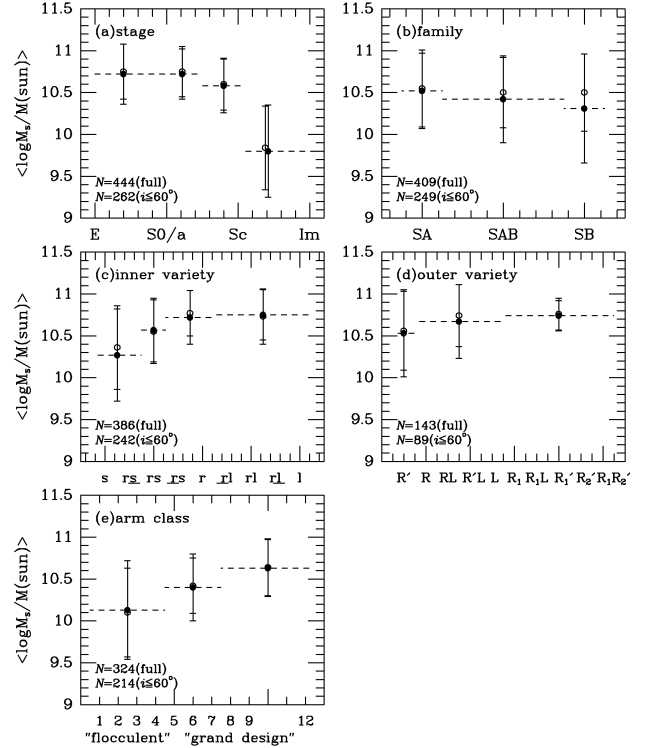}
\caption{Graphs of the mean log stellar mass of subsets of our AMIGA
sample versus morphological structure. Filled circles are for the full
(unrestricted by inclination) samples, while open circles are for the
subsets restricted to inclination $i$$\leq$60$^o$. The total numbers
of objects in each subset are indicated in each frame. The error bars
are 1$\sigma$ standard deviations about the means. The horizontal
dashed lines indicate the ranges of morphological features used in the
means.
}
\label{fig:stellmasses}
\end{figure}

Table~\ref{tab:smasses} summarizes the average stellar masses of the
isolated galaxies in our sample for which a mass estimate is available,
for the full samples and for the subsets restricted to $i$ $\leq$
60$^o$. Because we have mass estimates for only about 60\% of our
sample (Figure~\ref{fig:masses}), the log masses are averaged over a
range of types or morphological characteristics. The table highlights
some definite but not unexpected trends which are shown in
Figure~\ref{fig:stellmasses}. First, the early stages (E to Sab;
Figure~\ref{fig:stellmasses}a) are considerably more massive than the
later stages (S$\underline{\rm c}$d--Im), by a factor of nearly 8. The
intermediate stages (Sa$\underline{\rm b}$--Sc) are intermediate in
log mass but closer to the early types on average than the later
types. Second, the average log mass of the nonbarred (SA) galaxies in
the sample is slightly higher than that of the barred (SAB and SB)
galaxies (Figure~\ref{fig:stellmasses}b), although the effect is
smaller for the restricted subsets.

Among inner varieties, the isolated galaxies of types (s) and
(r$\underline{\rm s}$) are less massive than those of types
$\underline{\rm r}$s) and (r) by a factor of nearly 3, with (rs)
galaxies intermediate (Figure~\ref{fig:stellmasses}c). Galaxies of
types ($\underline{\rm r}$l), (rl), (r$\underline{\rm l}$), and (l)
have stellar masses comparable to those of types ($\underline{\rm
r}$s), and (r). Among outer varieties, isolated galaxies of types
(R$^{\prime}$) are less massive on average than all of the other outer
feature types, by a factor of 1.6 for both the full and restricted
subsets (Figure~\ref{fig:stellmasses}d). The highest outer variety
average mass is found for the outer resonant subclasses (R$_1$),
(R$_1^{\prime}$, R$_2^{\prime}$, etc.).

Figure~\ref{fig:stellmasses}e shows a tendency for flocculent spirals
(AC 1-4) to be less massive than grand design spirals (AC 8,9,12), by a
factor of 3-3.5. This is not surprising since the sophistication of
galactic structure is expected to be stronger for massive galaxies than
for low mass galaxies.

All of the trends shown in Figure~\ref{fig:stellmasses} can be
ultimately traced to the fact that later types are less luminous than
earlier types in general (e.g., de Vaucouleurs 1977; Buta. Corwin, and
Odewahn 2007). SB galaxies are more common at later types, inner rings
are more common at earlier types, outer pseudorings are more common at
later types than other kinds of outer features, and flocculent spirals
occur in later types as well.

\section{Comparative Histogram Analysis}

In this section, we use the Kolmogorov-Smirnov test to check for
potential morphological biases in Table~\ref{tab:catalog} due to
inclination and distance, to look for physical correlations with the
environment (Verley et al. 2007b), and to examine the distribution of
far-infrared (FIR) excess for bars, rings, and arm classes
(Figures~\ref{fig:adamas-1}-~\ref{fig:adamas-6b}), In each correlation
plot, we show the different morphological $T$-types, families, and
inner and outer varieties covered by the histograms.

The p-value of the Kolmogorov-Smirnov test is at the top of each plot
and was chosen for our analysis because it effectively quantifies how
much two distributions are independent of each other. In the case of a
three distribution comparison, we always show the KS test between the
two distributions that have the minimum p-value. We consider that two
distributions are significantly different when the p-value is lower
than 5\%. The $p$ values can be sensitive to the number of bins
used and to the number of elements involved in the comparison,
the result being more accurate when more elements are used.

\subsection{Stage, Family, Varieties, and Arm Class}

\begin{figure}
\vspace{-5truemm}
\includegraphics[width=\columnwidth]{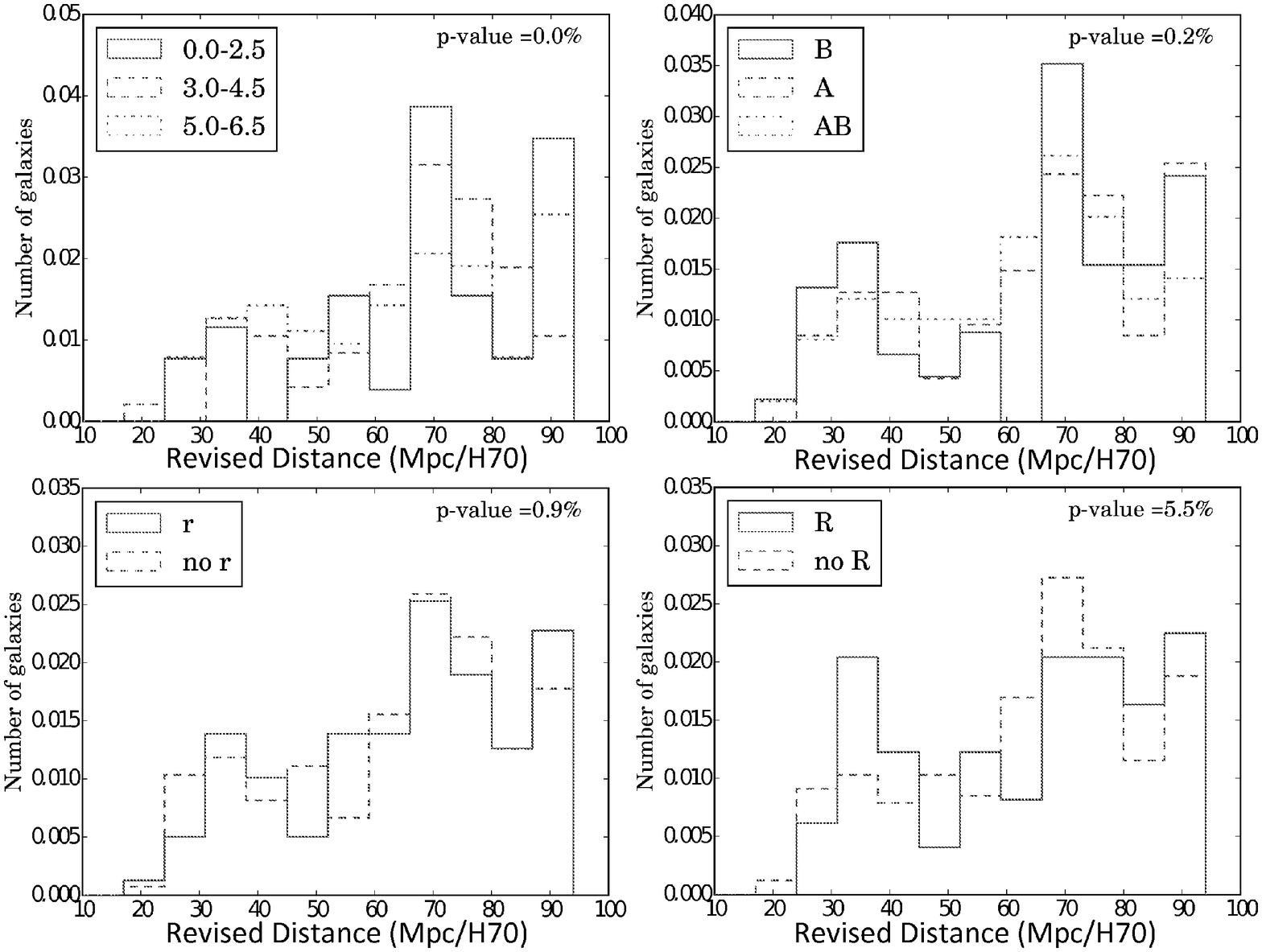}
\caption{Comparison between the relative frequency of stages, families,
inner varieties, and outer varieties in Table~\ref{tab:catalog} as a
function of distance $D$ in Mpc. The distances are derived from the
observed radial velocities assuming a Hubble constant of $H_0$ = 70 km
s$^{-1}$ Mpc$^{-1}$.}
\label{fig:adamas-1}
\end{figure}

\begin{figure}
\vspace{-5truemm}
\includegraphics[width=\columnwidth]{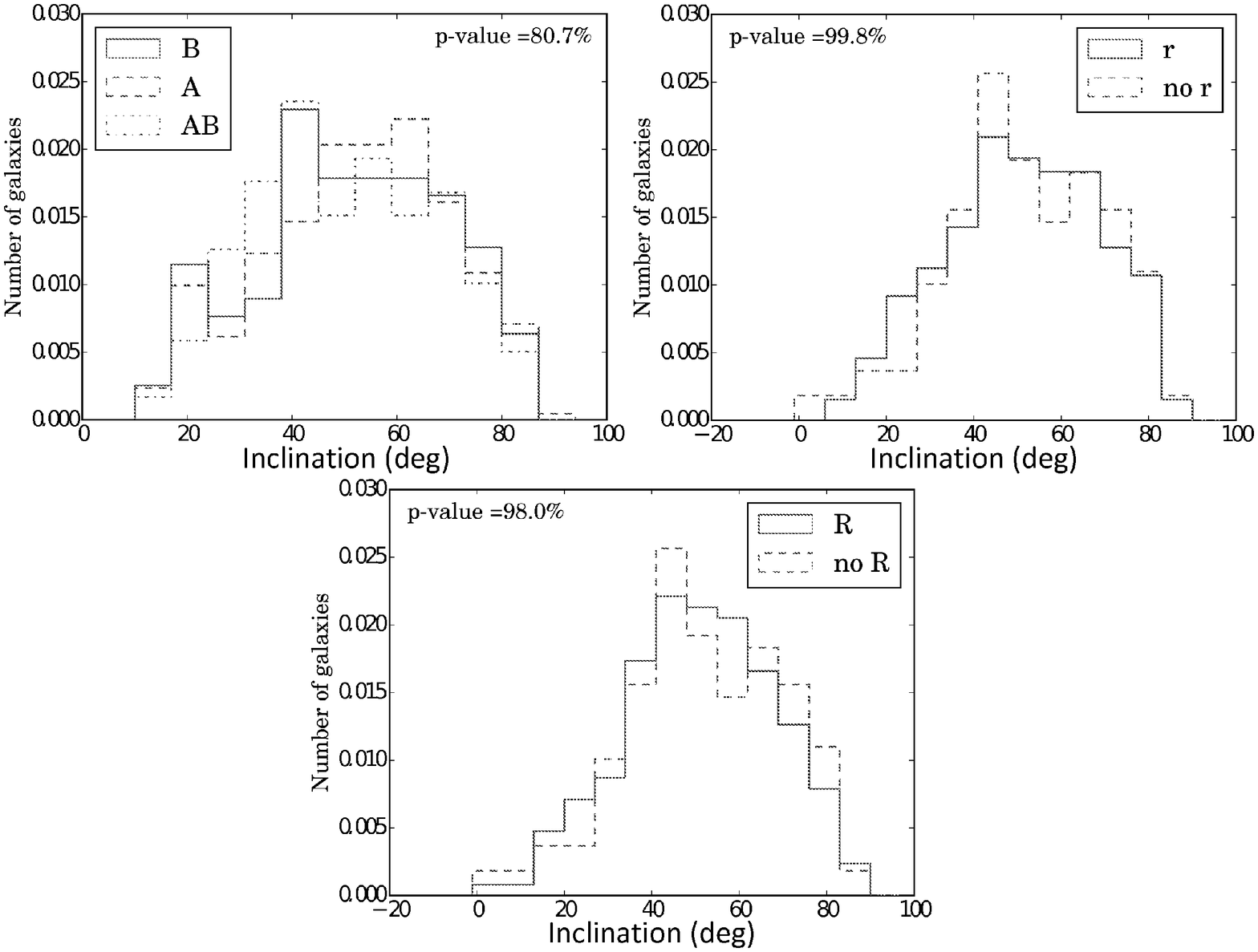}
\caption{Comparison between the relative frequency of families,
inner varieties, and outer varieties in Table~\ref{tab:catalog} as a
function of galaxy inclination.}
\label{fig:adamas-2b}
\end{figure}

First we look further for bias in the CVRHS classifications by
assessing whether the Table~\ref{tab:catalog} classifications are
significantly affected by either the source inclination or distance.
For example, if bars are more readily recognized in nearby galaxies
(potentially due to resolution limitations), then the distributions of
barred and non-barred galaxies with distance should be different, with
barred galaxies forming a distribution that favours smaller distances.
Figure~\ref{fig:adamas-1} compares the distributions of inner rings
(r), outer rings (R), bars (B, AB), and stages with distance (using
distances as calculated in Jones et al. 2018);
Figure~\ref{fig:adamas-2b} shows similar plots versus galaxy
inclination.

The KS test shows a difference between distributions (a p-value of
0.21\%) only in the case of the distance distribution versus
barred/non-barred galaxies (Figure~\ref{fig:adamas-1}, upper right
panel). Due to the relatively small range in redshift across the
sample, this difference is not likely due to evolution. In the case of
inner (r) and outer (R) rings, a small dependency is found (a p-value
of 5.46\%). Both galaxies with outer and inner rings in the isolated
sample are on average found at slightly larger distances than those
without these features. 

We repeated the tests above to look for bias due to galaxy inclination
(Figure~\ref{fig:adamas-2b}), but in all cases we found no apparent
dependency, consistent with what we illustrated in
Figures~\ref{fig:histos}-~\ref{fig:pvar}. However, whether or not a
spiral galaxy is identified as either grand design or flocculent is
clearly dependent on both distance and inclination. This is illustrated
in the top two panels of Figure~\ref{fig:adamas-6b}. The sample grand
design spirals seem to be more easily identified at lower inclinations
and on average are more distant than the sample flocculent spirals. The
dependence on inclination may not be unexpected because spiral arm
contrast is lower the more inclined the disk galaxy. The dependence on
distance could reflect the fact that grand design spirals tend to occur
in more luminous systems than do flocculent spirals, and hence can be
recognized to greater distances (section 7).

\subsection{Morphological features and isolation}

Using the same approach as above, we analyzed whether the distributions
of two environmental parameters: $\eta_k$ and $Q_{Kar,p}$ (Verley et
al. 2007b), were altered by restricting to subsamples of specific
morphological features. $\eta_k$ is the logarithm of the local volume
density of neighbouring galaxies, based on the projected separation to
the $k$th neighbour and the distance estimate of the galaxy in question
(Verley et al. 2007b). Only neighbours that have diameters that are
0.25-4 times that of the target galaxy are considered, as larger
(smaller) galaxies are assumed to be foreground (background) objects.
However, this does leave open the possibility that some low-mass dwarf
companions might be neglected. Typically $k$=5, but in some cases the
available optical field is too small to identify 5 neighbours and a
lower value of $k$ is used instead. $Q_{Kar,p}$ is a logarithmic and
dimensionless measurement of the ratio of a galaxy's internal
gravitational binding force to external tidal forces, which assumes
that galaxy diameter is a proxy for total mass (Verley et al. 2007b).
For CIG galaxies the mean values of $\eta_k$ and $Q_{Kar,p}$ are 1.4 and
$-$2.7, and their standard deviations are 0.6 and 0.7, respectively.

\begin{figure}
\vspace{-5truemm}
\includegraphics[width=\columnwidth]{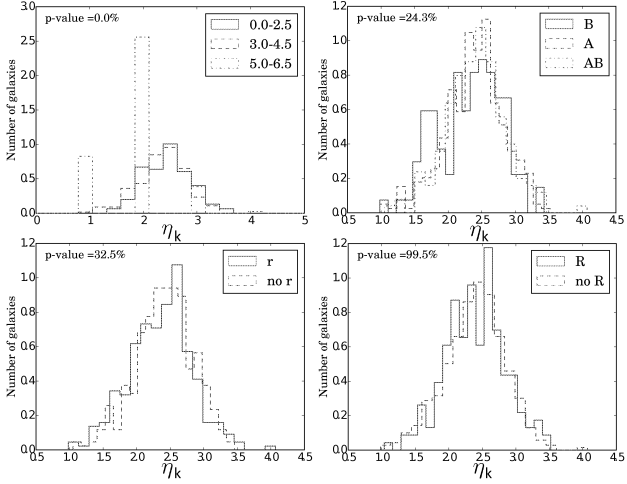}
\caption{Relative frequencies of stages, families, and inner and outer
varieties versus local density parameter $\eta_k$.}
\label{fig:adamas-4}
\end{figure}

\begin{figure}
\vspace{-5truemm}
\includegraphics[width=\columnwidth]{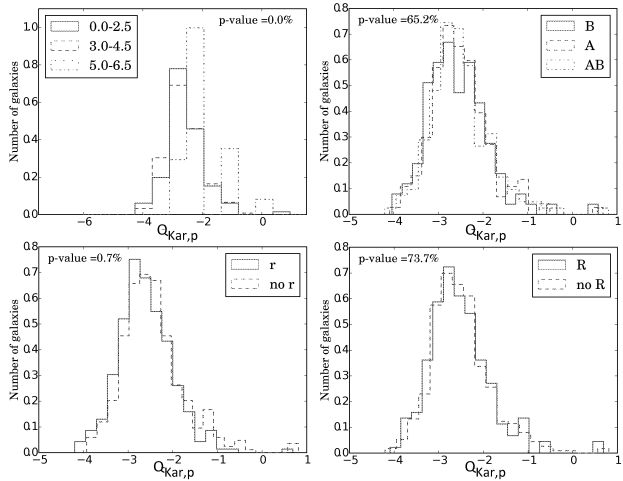}
\caption{Relative frequencies of stages, families, and inner and outer
varieties versus tidal strength parameter $Q_{Kar,p}$.}
\label{fig:adamas-3}
\end{figure}

Figures~\ref{fig:adamas-4}-~\ref{fig:adamas-3} show the comparative
histograms for stage, family, and inner and outer varieties versus
$\eta_k$ and $Q_{Kar,p}$; the comparisons for arm class are shown in
the middle two panels of Figure~\ref{fig:adamas-6b}. We find no
dependency on the isolation parameters regarding the detection of inner
or outer rings and bars, which could either be a sign of the minimal
impact of the environment on the generation of these features, or could
be indicating that the scatter in the measurement of the isolation
parameters is too large to allow us to distinguish between subtly
different environments which do and do not trigger bar/ring formation
in isolated galaxies. Another possibility is that isolated galaxies
which have bars/rings formed these features in past interactions,
although they would probably need to sustain those features for several
Gyr, as the AMIGA galaxies have been isolated on that timescale (see
section 1).

\subsection{Excess of FIR Luminosity}

The final parameter we examine is the correlation if any between the
morphological features recognized in the Table~\ref{tab:catalog} sample
and the FIR excess luminosity. Lisenfeld et al. (2007) derived scaling
relations for the AMIGA sample corresponding to $L_{FIR}$ versus $L_B$.
This fit gives the reference for isolated galaxies, although there is
some dispersion within the AMIGA sample itself which is likely due to
different degrees of isolation and different formation histories. It is
therefore worth checking if there is any correlation between some
structural features and a potential excess with respect to the
reference $L_{FIR}$, $L_B$ relation from that paper (that would
correspond to a star formation excess).

\begin{figure}
\vspace{-5truemm}
\includegraphics[width=\columnwidth]{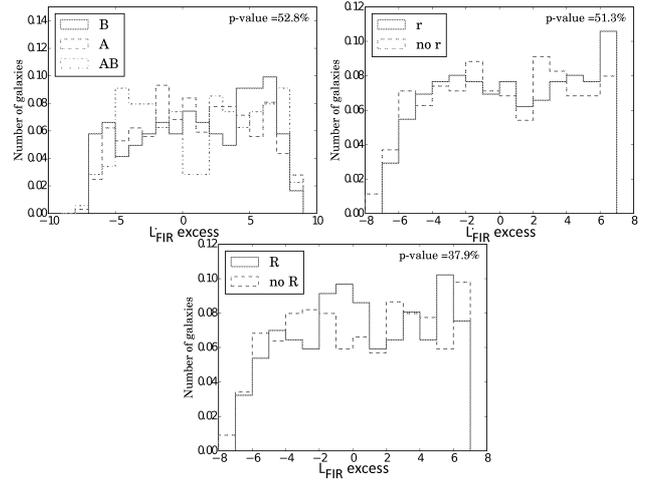}
\caption{Relative frequencies of stages, families, and inner and outer
varieties versus the FIR excess luminosity.}
\label{fig:adamas-5b}
\end{figure}

\begin{figure}
\vspace{-5truemm}
\includegraphics[width=\columnwidth]{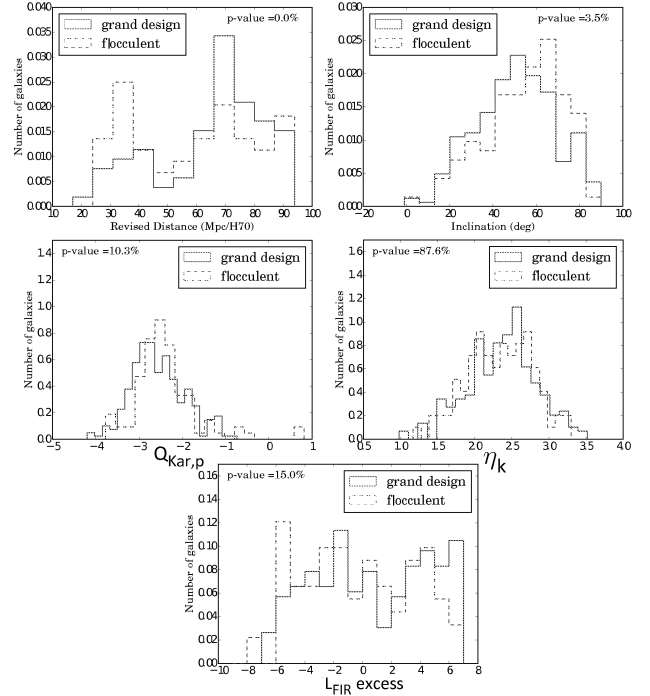}
\caption{Relative frequencies of 
arm classes versus distance, inclination, local density parameter $\eta_k$, 
tidal strength parameter $Q_{Kar,p}$, and FIR excess luminosity.}
\label{fig:adamas-6b}
\end{figure}

Figure~\ref{fig:adamas-5b} shows
the distributions for families and inner and outer varieties, while
Figure~\ref{fig:adamas-6b} (bottom panel) shows the distribution for
arm classes. This FIR deviation is measured according to the best-fit
(Lisenfeld et al. 2007) which was estimated by comparing the FIR
luminosity and the $B$-band luminosity of the AMIGA sample. Neither
bars, rings, nor arm classes for our Table~\ref{tab:catalog} sample
show a significant p-value from the KS test.

\section{Future Studies of Isolated Galaxies}

The type of study outlined in this paper depends strongly on the
quality of the images used for the classifications. SDSS images have
been very useful for this purpose, but new surveys have not only
improved on the SDSS, they also cover complementary parts of the sky.
For example, the Dark Energy Survey (DES, Abbott et al. 2019) is based
on the use of a specially designed wide field camera attached to the
Blanco 4-m telescope in Chile and covers 5000 square degrees of mostly
southern sky. The Kilo-Degree Survey (KiDS, de Jong et al. 2013) uses
the VLT Survey Telescope to cover 1500 square degrees of sky. The Hyper
Suprime-Cam Subaru Strategy Program (HSC-SSP, Aihara et al. 2018) and
the Large Synoptic Survey Telescope (LSST, Ivezi\`c et al. 2017) are
additional future sources of high quality imaging.

Although the goals of these new surveys are mainly cosmology-related
(with focus on dark energy and dark matter), the public availability of
these imaging databases will facilitate deeper, higher resolution
studies of thousands of nearby galaxies. These should allow improved
detection in some AMIGA galaxies of faint outer structures, like outer
rings or previously unrecognized low surface brightness features,
improved resolution of central structures, and also for the improved
visibility of very faint and possible small companions in the vicinity
of isolated galaxies. There must also be isolated galaxies in parts of
the sky that are not covered by the CIG that new surveys will
facilitate.

One of the values of the Table~\ref{tab:catalog} classifications is as
a training set for automatic classifications of faint galaxies that
will be present in the hundreds of thousands to millions on the imaging
of these new surveys. Deep machine learning techniques have been shown
to be very effective for such a purpose (e.g., Dom\'inguez S\'anchez et
al. 2018; Dieleman et al. 2015), and Table~\ref{tab:catalog} provides
an internally consistent set of classifications that could further
facilitate such studies.

\section{Summary}

We haved carried out a revised classification of 719 AMIGA candidate
isolated galaxies from the catalogue of Karachentseva (1973), based on
a set of $g$-band digital images from SDSS DR8. The classifications are
in the CVRHS system described by Buta et al. (2015). Our main findings
are:

\noindent
1. Consistent with previous studies, spirals are the dominant morphologies,
constituting nearly 85\% of the sample. Of these, the dominant subtypes are
Sb to Sc spirals.

\noindent
2. Visually strong bars have a low abundance in the AMIGA sample,
occurring at the 16\% level. Nonbarred galaxies, in contrast, make up
$\approx$50\% of the sample.

\noindent
3. (s)-variety spirals (i.e., spirals lacking an inner ring or
pseudoring) are the most abundant inner variety subtype, while no
outer feature is the most abundant outer variety subtype. In those
cases which do have an outer feature, outer pseudorings are the most
abundant outer variety subtype. These are not unusual characteristics
of a sample dominated by spirals.

\noindent
4. Grand design spirals are much more abundant in our isolated sample
than are flocculent spirals. However, as we have noted, this morphological
characteristic could only be reliably judged for 514 of the 597 spiral
galaxies recognized in the sample.

\noindent
5. S$\underline{\rm c}$d and later type galaxies in our sample are
less massive than E -- Sab galaxies by a factor of nearly 8. This
explains why (s)-variety spirals are less massive on average than
(r)-variety spirals, SB galaxies are slightly less massive on average
than SA galaxies, galaxies with outer rings, lenses, and resonant
subclasses are on average more massive than galaxies with ordinary
outer pseudorings R$^{\prime}$, and why ``grand design" spirals are
generally more massive than flocculent spirals.

\noindent
6. A comparative analysis of the distributions of morphological
features of isolated galaxies with distance, inclination, a local
density parameter, a tidal strength parameter, and Elmegreen arm class
reveals few significant correlations, based on the KS test. There may
be a slight distance bias in the recognition of bars in the sample, and
there may be both inclination and distance biases in the recognition of
arm classes.


We thank an anonymous reviewer for many helpful suggestions which
greatly improved this paper. We acknowledge support from grant
AYA2015-65973-C3-1-R (MINECO/FEDER, UE). This work has also been
supported by the Spanish Science Ministry ``Centro de Excelencia Severo
Ochoa" Program under grant SEV--2017-0709. M. Jones is supported by a
Juan de la Cierva formaci\'{o}n fellowship.

\end{document}